\documentclass{article}

\usepackage{graphicx}%
\usepackage{multirow}%
\usepackage{amsmath,amssymb,amsfonts}%
\usepackage{amsthm}%
\usepackage{mathrsfs}%
\usepackage[title]{appendix}%
\usepackage{xcolor}%
\usepackage{textcomp}%
\usepackage{manyfoot}%
\usepackage{enumitem}%
\usepackage{booktabs}%
\usepackage{algorithm}%
\usepackage{algorithmicx}%
\usepackage{algpseudocode}%
\usepackage{listings}%
\usepackage{bm}
\usepackage{caption} 

\usepackage{tabularx}
\usepackage{url}

\usepackage[authoryear]{natbib}

\usepackage{authblk}

\usepackage[margin=1.2in]{geometry}

\usepackage[colorlinks,citecolor=blue,urlcolor=blue]{hyperref}
\usepackage{doi}

\newcommand{\rcode}[1]{\normalfont\texttt{#1}}

\newcommand{\myheading}[1]{\par\medskip\noindent\textit{#1}\par\medskip\noindent\ignorespaces}

\newbox{\myorcidaffilbox}
\sbox{\myorcidaffilbox}{\large\includegraphics[height=1.7ex]{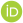}}
\newcommand{\orcidaffil}[1]{%
  \href{https://orcid.org/#1}{\usebox{\myorcidaffilbox}}}

\title{Unity Forests: Improving Interaction Modelling and Interpretability in Random Forests}
\author[1,2,*]{Roman Hornung  \orcidaffil{0000-0002-6036-1495}}
\author[3,4]{Alexander Hapfelmeier \orcidaffil{0000-0001-6765-6352}}
\affil[1]{Institute for Medical Information Processing, Biometry and Epidemiology, Faculty of Medicine, Ludwig Maximilian University of Munich (LMU), Munich, Germany}
\affil[2]{Munich Center for Machine Learning (MCML), Munich, Germany}
\affil[3]{Institute of General Practice and Health Services Research, Department Clinical Medicine, TUM School of Medicine and Health, Technical University of Munich (TUM), Munich, Germany}
\affil[4]{Institute of AI and Informatics in Medicine, TUM School of Medicine and Health, Technical University of Munich (TUM), Munich, Germany}
\affil[*]{Corresponding author: Roman Hornung, hornung@ibe.med.uni-muenchen.de}

\begin{document}

\maketitle

\begin{abstract}
Random forests (RFs) are widely used for prediction and variable importance analysis and are often believed to capture any types of interactions via recursive splitting. However, since the splits are chosen locally, interactions are only reliably captured when at least one involved covariate has a marginal effect. We introduce unity forests (UFOs), an RF variant designed to better exploit interactions involving covariates without marginal effects. In UFOs, the first few splits of each tree are optimized jointly across a random covariate subset to form a \lq\lq tree root'' capturing such interactions; the remainder is grown conventionally. We further propose the unity variable importance measure (VIM), which is based on out-of-bag split criterion values from the tree roots. Here, only a small fraction of tree root splits with the highest in-bag criterion values are considered per covariate, reflecting that covariates with purely interaction-based effects are discriminative only if a split in an interacting covariate occurred earlier in the tree. Finally, we introduce covariate-representative tree roots (CRTRs), which select representative tree roots per covariate and provide interpretable insight into the conditions - marginal or interactive - under which each covariate has its strongest effects. In a simulation study, the unity VIM reliably identified interacting covariates without marginal effects, unlike conventional RF-based VIMs. In a large-scale real-data comparison, UFOs achieved higher discrimination and predictive accuracy than standard RFs, with comparable calibration. The CRTRs reproduced the covariates’ true effect types reliably in simulated data and provided interesting insights in a real data analysis.
\end{abstract}

\section{Introduction}
\label{sec:introduction}

Random forests (RFs) \citep{Breiman:2001} are among the most widely used covariate-based prediction methods. Important reasons for the popularity of RFs include their generally strong predictive performance and the ability to rank covariates by predictive relevance using variable importance measures (VIMs). Among these, the permutation-based VIM \citep{Breiman:2001} is probably the most widely applied. For a long time, the mean decrease in impurity (MDI) \citep{Breiman:1984} was also popular, likely due to its lower computational cost. However, \citet{Strobl:2007} demonstrated that the MDI systematically assigns higher importance to covariates with many possible split points. To address this issue, bias-corrected versions of the MDI have been proposed, either based on permuted versions of the covariates \citep{Nembrini:2018} or by using the out-of-bag (OOB) data of the trees \citep{Li:2019, Loecher:2020, Zhou:2021}.

A key reason for the strong predictive performance of RFs lies in the low bias of their trees when modeling associations between the outcome and the covariates. In particular, RFs can capture interaction effects by recursively splitting the data into increasingly homogeneous subnodes using different covariates. However, because split selection is performed locally at each node without considering subsequent splits, only those interactions can be adequately represented for which at least one of the involved covariates exhibits a sufficiently strong marginal effect to be chosen as a split covariate \citep{Wright:2016, Hornung:2022}.

In this paper, we propose an alternative tree construction algorithm in which the first \textsc{max\_depth\_root} layers (default: 3) of each tree are determined jointly rather than sequentially. We refer to this algorithm as the \emph{Unity Forest} algorithm, and to the resulting forests as \emph{Unity Forests} (UFOs). The UFO algorithm permits splits that are not immediately discriminative---that is, splits for which the outcome distribution in the child nodes is no narrower than in the parent node---but which enable discriminative splits on other covariates in subsequent layers. Such combinations of splits can represent interaction effects in which none of the involved covariates exhibit a marginal effect. Although these purely interaction-based effects are usually less common than those involving marginal effects, they are often of substantive interest, and their influence cannot be captured by additive models such as generalized linear models without explicit specification.

In the following, we present several purely illustrative examples of interaction effects without marginal effects, illustrating how such effects can arise in applied contexts. Such effects are typically qualitative interaction effects, in which the direction of a covariate's influence depends on the value of another covariate. The direction of a therapy effect may depend on a moderator such as body mass index, with positive effects in individuals with high BMI and negative effects in those with low BMI. Psychological interventions may be beneficial in one contextual setting, such as a calm work environment, but counterproductive in another, such as a highly stressful setting. In agronomy, a fertilizer may increase plant growth on one soil type while inhibiting it on another due to differences in chemical or biological soil properties. Teaching methods such as case-based learning can enhance outcomes for students with high prior knowledge but impair learning for those with little prior knowledge. Similarly, interface elements like color-coded buttons may improve usability for younger users while reducing it for older users, for example due to age-related changes in color perception or familiarity with digital conventions.

In addition to the UFO algorithm, we propose a novel VIM, the \emph{unity VIM}. This measure aims to quantify the strength of each covariate’s influence under the conditions in which that influence is strongest. Here and throughout, the term \lq\lq influence'' is used for simplicity to refer to observed associations with the outcome, without making claims about causality. For a covariate that does not interact with others, there are no specific conditions under which it has a particularly strong effect, and the unity VIM reflects the strength of its marginal influence on the outcome. For covariates that interact with others, the relevant conditions correspond to subgroups defined by ranges in the values of those interacting covariates. If a covariate interacts with several others, only the subgroups involving its strongest interactions are considered. This ensures that only pronounced interaction effects contribute meaningfully to the unity VIM. Conversely, a covariate that exhibits only weak interactions with several others is not expected to obtain a high unity VIM value---an intentional feature, as weak interactions are difficult to detect and rarely of practical relevance in applications seeking to characterize the nature of covariate effects. Moreover, it is generally unlikely that a covariate engaging in multiple interactions would lack a direct effect on the outcome altogether.

As described in the previous paragraph, the unity VIM is designed to identify covariates that exhibit either a relevant marginal influence or conditional influence arising from interactions. However, the unity VIM values indicate only \emph{which} covariates are influential---not \emph{how} they affect the outcome. The latter aspect is particularly crucial for covariates whose effects are purely interaction-based. To better understand how such covariates influence the outcome, we propose a complementary method that aims to select, for each covariate, one tree that best represents its strongest effect. This procedure identifies a representative tree \citep{Laabs:2024}, the Covariate-Representative Tree Root (CRTR), for each covariate from the subset of trees in which the covariate shows its strongest OOB influence.

The complete UFO framework---comprising the UFO algorithm, the unity VIM, and the CRTR selection and visualization procedures---is implemented in the R package \rcode{unityForest}, available on CRAN. The package builds closely upon the widely used RF implementation \rcode{ranger} \citep{Wright:2017} and performs all computationally intensive operations in fast C++ code that uses all available CPU cores by default.

The remainder of this paper is organized as follows. Section~\ref{sec:rel_work} reviews related approaches. Section~\ref{sec:ufo_framework} describes the algorithms constituting the UFO framework. Section~\ref{sec:realdata_study} presents a large-scale real-data study in which we empirically evaluate key hyperparameters and compare the predictive performance of UFOs with that of conventional RFs. Section~\ref{sec:sim} reports a simulation study assessing the ability of the unity VIM to distinguish informative covariates, with a particular focus on interacting covariates, from non-informative ones relative to conventional VIMs. In Section~\ref{sec:crtr_analysis}, we evaluate the CRTR algorithm using both simulated and real data. Finally, Section~\ref{sec:discussion} discusses the results, limitations, and potential extensions, and Section~\ref{sec:sum_concl} summarizes the main findings.

\section{Related work}
\label{sec:rel_work}

UFOs are an RF variant with the goal of improving the structures of the trees in standard forests. Several approaches have aimed to include better performing trees in tree ensembles, but these usually rely on the conventional CART procedure \citep{Breiman:1984} for tree construction. For example, \citet{Bernard:2009} and \citet{Khan:2020} reduce the size of ensembles while seeking to improve predictive performance and interpretability, selecting subsets of trees based on performance criteria. An overview of related methods can be found in \citet{Manzali:2023}. In contrast, the UFO algorithm does not aim to select the best conventional trees, but to generate alternative tree structures that more holistically reflect covariate-outcome dependencies, in particular interaction effects without marginal effects. Its two main goals are: (i) leveraging these improved tree structures to enable a variable importance that better captures such interactions, and (ii) improving interpretability of the resulting trees.

To our knowledge, the only practically usable implementation of a RF-based VIM specifically designed to better capture interaction effects is the Shapley-based approach SHAFF by \citet{Benard:2022}, which, however, is limited to continuous outcomes.

Two further areas of related work are rule extraction methods and approaches to select representative trees, both conceptually linked to the CRTR algorithm proposed in this paper. Rule extraction methods such as RuleFit \citep{Friedman:2008}, node harvest \citep{Meinshausen:2010}, and SIRUS \citep{Benard:2021} aim to summarize forests into sparse sets of decision rules, thereby enhancing interpretability while retaining predictive power. These methods differ in how they generate, truncate, and combine rules, but all share the idea of making the implicit structure of forests more transparent.

Approaches to representative trees aim to identify trees that are most characteristic of the ensemble for inspection. They define pairwise distances between trees and then select a representative tree as the one with the smallest average distance to all other trees. \citet{Banerjee:2012} introduced this idea, and \citet{Laabs:2024} refined it by weighting upper-level splits more strongly in the distance measure. While single trees highlight characteristic structures, they underrepresent the complexity of forests. To address this, \citet{Szepannek:2024} proposed using small sets of representative or surrogate trees, balancing interpretability with a more comprehensive coverage of the relations captured by the forest.

\section{The Unity Forest Framework}
\label{sec:ufo_framework}

Let $(\bm{x}_i, y_i)$ denote the $i$-th observation, where $\bm{x}_i = (x_{i1}, \ldots, x_{ip})^\top$ is a $p$-dimensional vector of covariates and $y_i$ the associated outcome. We consider classification and regression, where in the case of classification, $y_i \in \{1, \dots, K\}$ denotes a class label, whereas for regression, $y_i \in \mathbb{R}$ is a continuous outcome. For classification, we consider both class label prediction and class probability estimation. Class labels are predicted by majority voting, as in conventional RFs, whereas class probabilities are obtained by averaging the empirical class distributions in the leaf nodes containing the observations to be predicted, as described by \citet{Malley:2012}.

In this section, we describe the UFO algorithm for tree construction, the algorithm for calculating the unity VIM, and the algorithm for selecting CRTRs.

\subsection{Tree Construction}
\label{sec:tree_constr}

\subsubsection{Algorithm and Rationales}
\label{sec:ufo_tree_alg}

To provide initial orientation, Algorithm~\ref{alg:tree_constr} presents a high-level description of the procedure used to construct each individual tree in a UFO. A detailed step-by-step description, including specific stop rules and other technical details, can be found in Section A.1 of the supplementary material.

\begin{algorithm}[H]
\caption{Sketch of the tree construction procedure in unity forests}
\label{alg:tree_constr}
\begin{algorithmic}[1]
\State Subsample, without replacement, a proportion \textsc{fract\_n} (default: 0.7) of the observations---the \emph{tree sample}---and a smaller proportion \textsc{prop\_var} of the covariates. The default value for $\textsc{prop\_var}$ is set to $\sqrt{p}/p$ for $p \leq 100$ and to 0.1 for $p > 100$, where $p$ denotes the number of covariates.
\vspace{0.3cm}
\State Grow \textsc{n\_cand\_trees} (default: 500) candidate \emph{tree roots} of limited depth \textsc{max\_depth\_root} (default: 3) on the tree sample using the selected covariates, with both split covariates and split points chosen at random.
\vspace{0.3cm}
\State For each candidate tree root, calculate a \emph{partition criterion} that measures the reduction in impurity achieved by the full set of \emph{root leaves} (i.e., the leaf nodes of the tree roots) compared to the impurity of the entire tree sample---using Gini impurity for classification and variance for regression.
\vspace{0.3cm}
\State Select the candidate tree root with the best partition criterion value; since the impurity of the tree sample is constant, this corresponds to selecting the root with the purest set of root leaves.
\vspace{0.3cm}
\State Expand this tree root into a full tree using the tree sample and conventional recursive CART-style splitting, selecting random subsets of $\textsc{mtry}$ (default: $\lfloor \sqrt{p} \rfloor$) candidate covariates from all covariates at each split, as in conventional RFs.
\end{algorithmic}
\end{algorithm}

While the term \lq\lq random forests'' (RFs) today usually refers to a specific procedure, the definition in the original paper by \citet{Breiman:2001} is considerably more general. There, RFs are defined as a collection of trees in which each tree depends on a random vector~$\Theta$, independently and identically distributed across trees. Since Algorithm~\ref{alg:tree_constr} complies with this definition, UFOs constitute an RF variant.

In the UFO algorithm, the splits in the tree roots are not fully optimized. Instead, for each tree, the tree root that yields the purest root leaves is selected from many candidate roots. As a result, the chosen tree roots will generally not produce the best partitioning possible with the covariates randomly drawn in step~1 of Algorithm~\ref{alg:tree_constr}, although increasingly good approximations would be obtained if the number of candidate roots were increased. However, this is not required. The predictive performance of RFs tends to increase with better-performing individual trees but decreases as the predictions of different trees become more correlated \citep{Breiman:2001}. In conventional RFs, each split is determined from a random subset of $\textsc{mtry}$ covariates, which are resampled for every split, thereby ensuring high tree diversity. If, in contrast, the splits in the tree roots were optimized in the UFO algorithm, the correlation between tree predictions would increase, potentially reducing the overall predictive performance of the forest.

If the optimal partitioning were found, the resulting tree roots would correspond to so-called optimal trees \citep{Bertsimas:2017}, whose construction is feasible today but remains computationally demanding. Similar to optimal trees, the selection of the best tree roots in the UFO algorithm seeks the most effective partitioning of the covariate space spanned by the randomly selected covariates, rather than locally discriminative splits at individual nodes as in conventional CART trees. This approach allows the UFO algorithm to include covariates that may have no marginal effect themselves but that contribute to an effective partitioning through subsequent splits on other covariates, thereby more holistically capturing interaction effects between covariates.

Not all covariates are considered for constructing the tree roots. Only a subset is used, both to enhance tree diversity---similar in effect to subsampling observations and using non-optimized splits---and because this restriction plays an important role for the unity VIM, which is discussed in Section~\ref{sec:unity_vim}.

The tree roots are subsequently expanded into full trees because deep trees are associated with lower bias. Although deep trees exhibit higher variance, this variance does not propagate to the forest predictions, as the results are averaged over many trees \citep[p.~588]{Hastie:2009}.

\subsubsection{Handling of Nominal Covariates}
The tree construction procedure in the UFO framework does not directly extend to nominal covariates, as it requires ordered covariate values. To nevertheless include nominal covariates, we order their categories following the approach implemented in the R package \rcode{ranger} (version~0.17.0) using the option \rcode{respect.unordered.factors = "order"} \citep{Wright:2019, Coppersmith:1999}. The underlying idea of this option is to arrange the categories of a nominal covariate such that, when progressing along the order of categories, the corresponding outcome values tend to change in the same direction. The specific procedures applied for categorical and continuous outcomes are described in detail in Section~A.2 of the supplementary material. Based on the resulting order, the categories of each nominal covariate are assigned numeric values $1, \dots, J$, where $J$ denotes the number of categories, and are subsequently treated as ordered covariates within the UFO framework.

\subsubsection{Hyperparameters}
The UFO algorithm involves several hyperparameters. We recommend using the default values for these rather than performing optimization. This recommendation is based on two considerations: (i) hyperparameter tuning typically yields only minor improvements in the predictive performance of RFs \citep{Probst:2019}, and (ii) the optimal values of certain parameters depend on others, making isolated optimization potentially problematic.

Many of the hyperparameters correspond to those of conventional RFs, and their defaults closely follow the implementation in \rcode{ranger} \citep{Wright:2017}. In Section~\ref{sec:prestudy}, we assess the suitability of the default values for three critical hyperparameters using 30 real-world datasets.

A hyperparameter that affects interpretability is $\textsc{max\_depth\_root}$, which defaults to~3. This setting allows the tree roots to capture interaction effects involving up to three covariates.

A complete list of all hyperparameters, together with their default values and detailed explanations, is provided in Section~A.3 of the supplementary material.

\subsection{Unity VIM}
\label{sec:unity_vim}

Algorithm~\ref{alg:unity_vim} provides a high-level overview of the procedure for calculating the unity VIM values as an initial orientation.

\begin{algorithm}[H]
\caption{Sketch of the calculation of the unity VIM}
\label{alg:unity_vim}
\begin{algorithmic}[1]
\State For each node in the tree roots that is not a root leaf, using the in-bag observations, compute a \emph{split score} that measures the divisiveness of the corresponding split.
\vspace{0.3cm}
\State For each covariate, select the set of top-scoring splits.
\vspace{0.3cm}
\State For each covariate, at the top-scoring splits, calculate the unity VIM based on differences between the OOB-based split scores before and after permuting the covariate within the nodes.
\end{algorithmic}
\end{algorithm}

The unity VIM focuses on the most discriminative splits of each covariate, thereby achieving its intended purpose of quantifying covariate influence exclusively under the conditions in which that influence is strongest. For a covariate~A with a strong marginal effect, the vast majority of its splits will be discriminative. In contrast, for a covariate~B that exhibits little or no marginal effect but interacts with another covariate, splits in~B will be discriminative only if a prior split in the same tree has occurred in the covariate with which~B interacts. Consequently, most splits in~B will not be discriminative---an effect further amplified by the fact that only a small proportion~$\textsc{prop\_var}$ of the covariates (see Section~\ref{sec:ufo_tree_alg}) is considered for each tree root. For such covariates, it is therefore essential that only the most discriminative splits contribute to the unity VIM.

Apart from its role in increasing tree diversity (see Section~\ref{sec:ufo_tree_alg}), restricting each tree root to a small proportion of covariates also mitigates dominance by the same strongly influential covariates across the forest. This, in turn, allows interaction effects without marginal effects to be captured in many trees.

As outlined in Algorithm~\ref{alg:unity_vim}, the computation of split scores and the subsequent selection of the top splits are performed using the in-bag observations, whereas the contributions to the unity VIM based on the best splits are computed using the  OOB  data. Using distinct data subsets for split selection and for evaluating the discriminative power of the selected splits prevents an upward bias in the unity VIM values that would otherwise result from optimal selection on the same data.

The detailed algorithm for calculating the unity VIM proceeds as follows:

\begin{enumerate}
\item \emph{Compute split scores quantifying split discriminativeness.}
Using the in-bag observations, compute for each split at the internal nodes of the tree roots (i.e., excluding the root leaves) across all trees the reduction in Gini impurity (for classification) or in variance (for regression), and multiply this quantity by the node size. Denote the resulting values as \emph{split scores}.
\item \emph{Select top-scoring splits per covariate.}
For each covariate, identify the proportion $\textsc{prop\_best\_splits}$ (default: 0.01) of splits within the tree roots that have the largest split scores among all splits in the tree roots using that covariate. If this proportion corresponds to fewer than five splits, select the five highest-scoring splits or all available splits if fewer than five exist for that covariate. Denote the indices of the nodes corresponding to the selected splits for covariate~$j$ by~$\mathcal{B}_j$.
\item \emph{Compute the unity VIM using OOB-based split scores at the top-scoring splits.}
For each covariate~$j$, $j \in {1, \dots, p}$, calculate the unity VIM as:
\begin{align}
\sum_{l \in \mathcal{B}_j} N_l (\text{OOB\_SC}_l - \text{OOB\_SC\_PERM}_l).
\end{align}
Here, $N_l$ denotes the number of in-bag observations in node~$l$.
$\text{OOB\_SC}_l$ is the reduction in Gini impurity (for classification) or variance (for regression) at node~$l$, computed using only the OOB observations that pass through this node.
$\text{OOB\_SC\_PERM}_l$ is the same quantity, but calculated after randomly permuting the values of covariate~$j$ among the OOB observations that pass node~$l$.
\end{enumerate}

The fallback to selecting (up to) five splits is included to stabilize the unity VIM in cases where the proportion-based rule (\textsc{prop\_best\_splits}) would yield fewer than five splits for a covariate. Basing the VIM on only too small a number of splits would make it more variable and sensitive to random fluctuations. The lower bound of five therefore ensures that the unity VIM is computed from a sufficiently large set of informative splits, while still focusing on the most discriminative ones. In our simulation study and real-data analyses, this fallback was never triggered.

Further explanations of the algorithm and discussions of the default hyperparameter values are provided in Section~A.4 of the supplementary material.

\subsection{Covariate-Representative Tree Roots (CRTRs)}
\label{sec:crtrs}

As already mentioned, CRTRs can, in principle, be selected for any covariate. However, since only covariates with adequately high unity VIM values can be assumed to have a relevant influence---either marginally or through interactions with other covariates---it is advisable to select CRTRs only for such covariates.

The algorithm for selecting CRTRs is described in detail below:

\begin{enumerate}
\item \emph{Compute the split scores using the OOB data.}
Compute split scores for each internal node of the tree roots in the same manner as in the unity VIM calculation, but use the OOB observations instead of the in-bag observations.
\item \emph{Select the top-scoring splits per covariate.}
Perform this step exactly as in the unity VIM calculation, using the same proportion $\textsc{prop\_best\_splits}$.
\item \emph{Select the best tree roots.}
For each covariate, identify the set of tree roots that contain at least one top-scoring split for that covariate. Denote the selected set of tree roots as the \emph{best tree roots} of the corresponding covariate.
\item \emph{Select the CRTRs.}
For each covariate, determine the CRTR from the set of best tree roots using the procedure of \citet{Laabs:2024}.
\item \emph{Assign covariate scores for visualization.}
In the selected CRTRs, assign to the internal nodes of the tree roots a \emph{covariate score} calculated as:
\begin{equation}
\frac{\#{freq\_best\_j}}{\#{freq\_best\_j} + \#{freq\_all\_j}},
\end{equation}
where covariate $j$ is the split covariate at the respective node. Here, $\#{freq\_best\_j}$ denotes the relative frequency with which covariate~$j$ is used for splits at the internal nodes of the best tree roots for covariate~$j$, and $\#{freq\_all\_j}$ denotes the corresponding relative frequency of splits on covariate~$j$ across the internal nodes of all tree roots. These covariate scores are used for visualizing the representative tree roots; see Section~\ref{sec:crtr_sim} for details.
\end{enumerate}

By selecting the tree roots in which the respective covariate induces the most discriminative split or splits, the resulting best tree roots primarily represent conditions under which the covariate exerts its strongest influence. Naturally, some variability remains within these sets, as not all tree roots capture these conditions equally well. The approach of \citet{Laabs:2024} selects the tree root with the smallest average distance to all other tree roots in the set of best tree roots, based on a distance measure reflecting the similarity of covariate usage. Splits at higher levels are weighted more heavily than those at lower levels, as the former have a greater impact on the tree structure and the resulting partitioning of the covariate space, making it unlikely that the selected tree root will insufficiently reflect the conditions under which the covariate has the strongest effect.

Unlike in the calculation of the unity VIM, in the above algorithm the top-scoring splits---and thus the best tree roots---are selected using the OOB data rather than the in-bag data. The rationale is that, during CRTR visualization, the node-level outcome distributions or summary statistics used for interpretation purposes are computed from the in-bag data. If the selection of the best tree roots were based on the same in-bag data, the splits based on the CRTR covariates would exhibit artificially high discriminatory power due to optimal selection. Using the OOB data for selection avoids this optimistic bias because the in-bag-based annotations in the visualization remain unbiased.

In Section~\ref{sec:crtr_analysis}, we present applications of CRTRs to simulated and real data and describe the visualization procedure in detail.

\section{Large-Scale Real-Data Evaluation of the Predictive Performance of UFOs}
\label{sec:realdata_study}

In this section, we evaluate the predictive performance of UFOs on a large number of real-world datasets with binary outcomes (see Section~\ref{sec:discussion} for a discussion of other outcome types). Section~\ref{sec:data} describes the dataset pool used in this study. In Section~\ref{sec:prestudy}, we conduct a preliminary analysis to assess the suitability of the default values chosen for several critical hyperparameters using a relatively small subset of the datasets. Finally, Section~\ref{sec:compstudy} presents a large-scale comparison of the predictive performance of UFOs and conventional RFs based on the remaining datasets.

\subsection{Data}
\label{sec:data}

We used 198 real datasets with binary outcomes. These datasets form a preprocessed subset of the 243 datasets used in the large-scale benchmark study by \citet{Couronne:2018}. We excluded redundant, highly similar, or statistically dependent datasets to minimize dependencies across the data source. To limit computational demands, datasets with more than 10{,}000 observations or more than 1{,}000 covariates were randomly subsetted to contain 10{,}000 observations or 1{,}000 covariates, respectively. An overview of the datasets used is provided in Tables~S1–S5 of the supplementary material.

To ensure full reproducibility, the R code and data required for all analyses presented in this paper and in the supplementary material are provided on GitHub (\url{https://github.com/RomanHornung/UnityForests_code_and_data}, commit: 08edff9). Here, the reasons for the exclusion of 45 datasets from the original 243 are described in detailed in a dedicated R script.

\subsection{Preliminary Study on the Appropriateness of Critical Default Hyperparameter Values}
\label{sec:prestudy}

The UFO framework involves several hyperparameters. In Section~\ref{sec:ufo_framework} and in Sections~A.3 and~A.4 of the supplementary material, we specified and justified the chosen default values. While, as previously noted, RF performance is generally quite robust to changes in the default hyperparameter values, we conducted a targeted evaluation of three hyperparameters that could plausibly have a stronger influence on predictive performance: the number of candidate trees ($\textsc{n\_cand\_trees}$), the depth of tree roots ($\textsc{max\_depth\_root}$), and the proportion of covariates sampled per tree root ($\textsc{prop\_var}$).

The aim of this preliminary analysis was not to identify jointly optimal hyperparameter combinations, but to assess whether the chosen default values yield predictive performance close to the best performance attainable within reasonable ranges and to evaluate how sensitive predictive performance is to changes in each hyperparameter. Moreover, the default values were chosen based on conceptual considerations, and the analysis was not intended to override these design choices.

For this preliminary analysis, we used a subset of 30 datasets from the total of 198. Each of the three hyperparameters was varied separately across a grid of reasonable values, including the respective default value, while all other hyperparameters were kept fixed at their default values. The predictive performance of UFOs was assessed using repeated five-fold cross-validation with three performance measures: the Brier score, the area under the ROC curve (AUC), and the accuracy. Section~B.2 of the supplementary material describes the study design and the obtained results in detail.

In summary, the examined hyperparameters had a negligible impact on predictive performance. For the vast majority of datasets, no notable performance differences were observed across the values considered. In cases where small differences across the considered hyperparameter values occurred, the performance achieved with the default values was equal to or very close to the maximum observed performance. These findings indicate that the selected default values for the three hyperparameters are appropriate and that tuning these parameters is unlikely to yield meaningful performance improvements in most applications.

\subsection{Comparison of Predictive Performance between UFOs and Conventional RFs}
\label{sec:compstudy}

As UFOs represent a modification of conventional RFs---a widely used and well-established method whose predictive performance has been extensively studied---it is of particular interest to assess whether this modification leads to improvements or deteriorations in predictive performance. To this end, we compared the predictive performance of UFOs and conventional RFs using 168 of the 198 available datasets, excluding those that were used in the preliminary study.

For UFOs, all hyperparameters were set to their default values. To ensure a fair comparison under comparable default settings and avoid favoring either method, we configured the conventional RFs to match the UFO setup as closely as possible, keeping all RF hyperparameters at the corresponding UFO default values. Specifically, we employed subsampling with $\textsc{fract\_n} = 0.7$ instead of bootstrapping, used 20{,}000 trees per forest, and set the minimum node size for splitting to five, consistent with the fact that we performed probability prediction. As in the preliminary study, we applied five-fold stratified cross-validation repeated five times as the resampling scheme, and we used the Brier score (\rcode{Brier}), the area under the ROC curve (\rcode{AUC}), and the accuracy (\rcode{ACC}) as performance metrics.

Table~\ref{tab:benchmark_results} summarizes, on a dataset-wise basis, how often UFOs performed better, equally well (i.e., identical performance values), or worse than conventional RFs across datasets for the different performance measures. Regarding \rcode{Brier}, UFOs performed worse than RFs on slightly more than half of the datasets. In contrast, for \rcode{AUC}, UFOs outperformed RFs on roughly two-thirds of the datasets and performed worse on about one-quarter. UFOs also outperformed RFs in terms of \rcode{ACC}, performing better in approximately 60\% of cases and worse in about one-third.

\begin{table}[!ht]
\centering
\caption{Comparison of UFOs and conventional RFs across datasets. Each entry shows the number of datasets for which UFOs performed better, equal, or worse than RFs; percentages in parentheses indicate the corresponding proportions} 
\label{tab:benchmark_results}
\begin{tabular}{lccc}
  \toprule
Metric & UFO better & UFO equal & UFO worse \\ 
  \midrule
  Brier & 78 (46.4\%) & 0 (0.0\%) & 90 (53.6\%) \\ 
  AUC & 114 (67.9\%) & 9 (5.4\%) & 45 (26.8\%) \\ 
  ACC & 101 (60.1\%) & 10 (6.0\%) & 57 (33.9\%) \\ 
   \bottomrule
\end{tabular}
\end{table}

To assess whether UFOs performed statistically significantly better or worse than RFs, we conducted binomial tests, treating datasets as independent observations. The null hypothesis was that the probability of UFOs achieving higher predictive performance than RFs equals 0.5. For each test, datasets where both methods achieved identical results were excluded, as these cases provide no information about relative performance. The resulting $p$-values were as follows: \rcode{Brier}: 0.396; \rcode{AUC}: $< 0.001$; \rcode{ACC}: $< 0.001$. Using the conventional significance level $\alpha = 0.05$, we conclude that UFOs performed significantly better than RFs in terms of \rcode{AUC} and \rcode{ACC}, while the difference in \rcode{Brier} was not statistically significant.

In summary, UFOs demonstrated statistically significant improvements over RFs with respect to \rcode{AUC} and \rcode{ACC}, whereas RFs were slightly, though not significantly, more frequently superior in terms of \rcode{Brier}. A supplementary analysis of the dataset-specific performance measure values (see Section~B.3 of the supplementary material) showed that, in the vast majority of cases, the differences between the two methods were small.

\section{Simulation Study Comparing the Unity VIM with Conventional VIMs}
\label{sec:sim}

The unity VIM was designed to identify both covariates with purely marginal effects and those that influence the outcome primarily through interactions with other covariates---where the interacting covariates themselves may or may not have marginal effects. Two central components of the UFO framework are particularly relevant in this context: 1) the simultaneous optimization of the splits in the tree roots, which prevents the underselection of interacting covariates without marginal effects, and 2) the inclusion of only the most discriminative splits for each covariate in the unity VIM calculation, ensuring that the effects of covariates acting only in specific subgroups defined by other covariates are adequately taken into account. 

In contrast, conventional RF-based VIMs are derived from standard trees, in which each split is chosen based solely on its local discriminatory power at the current node. Moreover, these VIMs aggregate information across all splits for each covariate, which may underestimate the importance of covariates that act only under specific interaction conditions.

Given these conceptual differences, the primary goal of the present simulation study was to compare the unity VIM (\rcode{Unity\_vim}) with two conventional RF-based VIMs---the permutation VIM (\rcode{Perm}) and the corrected Gini importance (\rcode{Gini})---in terms of their ability to distinguish informative covariates (with marginal or interaction effects) from non-informative covariates. Additionally, we investigated the extent to which covariates with different effect types are ranked differently by the three VIMs.

\subsection{Study Design}
\label{sec:sim_design}

The UFOs (unity VIM) and conventional RFs (permutation VIM, corrected Gini importance) used to compute the three compared VIMs were configured in the same way as in the real-data analysis described in Section~\ref{sec:compstudy}, ensuring comparability of results.

\subsubsection{Data-Generating Processes}

We considered two data-generating processes (DGPs), both with balanced binary outcomes. The first DGP included only continuous covariates, while the second comprised both continuous and categorical covariates.

For each DGP, we examined sample sizes of 100, 300, 500, and 1{,}000 and generated 1{,}000 datasets for each of these.

\myheading{Data-Generating Process 1: Continuous Covariates Only}
This DGP, hereafter referred to as DGP~1, was adopted from \citet{Hornung:2022}, who used it in a comparative study of RF-based approaches for ranking pairwise interaction effects. It includes 18 informative and 50 non-informative covariates.

\begin{figure}[!ht]
\centering
\includegraphics[width=0.95\linewidth]{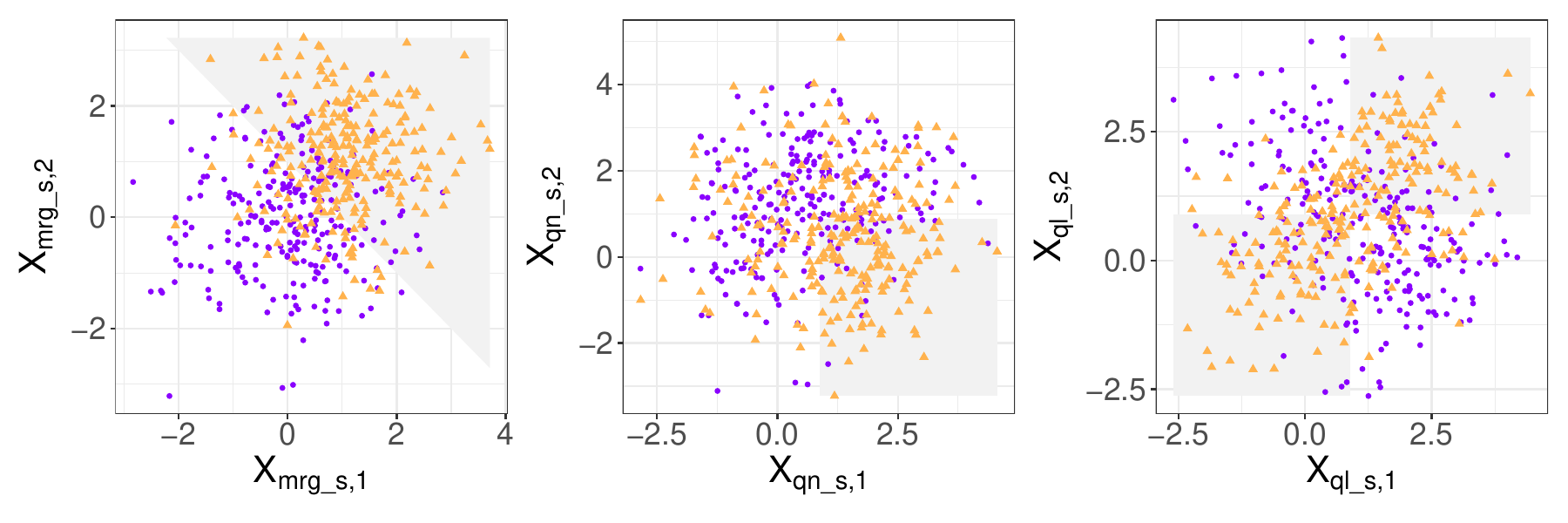}
\caption{Example pairs of variables with strong effects in a simulated dataset generated from DGP~1 ($n=500$). Circles and triangles represent observations from class 1 and 2, respectively. Shaded regions indicate areas of the covariate space with an above-average concentration of observations from class 2}
\label{fig:sim_des_1_example_data}
\end{figure}

Figure~\ref{fig:sim_des_1_example_data} illustrates the values of representative covariates for each type of informative effect in one simulated dataset. The left panel shows two covariates with purely marginal effects, the middle panel depicts a pair involved in a quantitative interaction (both covariates having marginal effects), and the right panel shows a pair exhibiting a qualitative interaction without marginal effects. Quantitative (bivariate) interactions refer to cases in which the strength but not the direction of one covariate's effect depends on the other covariate. In contrast, as previously mentioned, qualitative interactions involve both strength and direction changes. Section~C.1.1 of the supplementary material provides a detailed description of the effect mechanisms for the covariates shown in Figure~\ref{fig:sim_des_1_example_data}. 

The covariate types $X_{mrg\_s}$, $X_{qn\_s}$, and $X_{ql\_s}$ depicted in Figure~\ref{fig:sim_des_1_example_data} represent strong effects. For each of these, we also included variants with moderate and weak effect strengths in the DGP, featuring the same mechanisms but differing magnitudes of association. These variants are denoted by subscripts \lq\lq \_m'' (moderate) and \lq\lq \_w'' (weak), replacing \lq\lq \_s'' (strong) in the variable names.

The 18 informative covariates in DGP~1 consist of two covariates for each of the three purely marginal effect types and one pair for each of the six interaction types.

A detailed specification of this DGP is provided in Section~C.2.1 of the supplementary material.

\myheading{Data-Generating Process 2: Continuous and Categorical Covariates}
The second DGP, referred to as DGP~2, includes 19 informative covariates---four of which are categorical---and 50 continuous non-informative covariates.

\begin{figure}[!ht]
\centering
\includegraphics[width=0.95\linewidth]{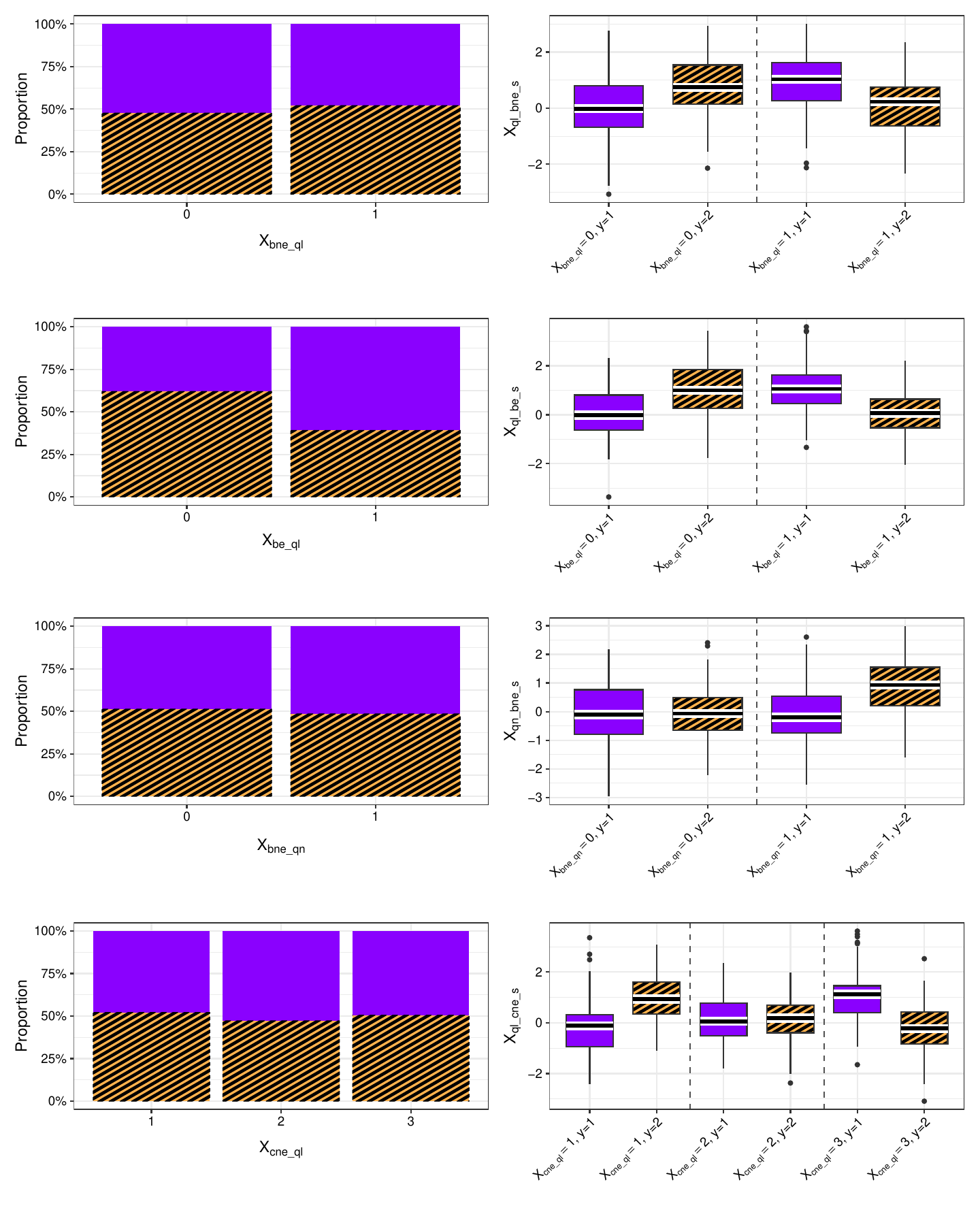}
\caption{Empirical effects of the interacting covariates in a simulated dataset generated from DGP~2 ($n=500$). In the bar plots, solid and hatched areas represent the proportions of classes 1 and 2, respectively. In the box plots, solid and hatched boxes correspond to the covariate values of classes 1 and 2, respectively}
\label{fig:sim_des_2_example_data}
\end{figure}

Three informative covariates, $X_{mrg\_s}$, $X_{mrg\_m}$, and $X_{mrg\_w}$, have only marginal effects identical to those in DGP~1 and are therefore not discussed further here. Figure~\ref{fig:sim_des_2_example_data} shows the empirical effects of all remaining informative covariates in one simulated dataset. Each row corresponds to a pairwise interaction between two covariates. Section~C.1.2 of the supplementary material provides a detailed explanation of these effect mechanisms.

As in DGP~1, we also included moderate and weak effect-size variants of the covariate types with strong effects shown in Figure~\ref{fig:sim_des_2_example_data}. Specifically, in addition to $X_{ql\_bne\_s}$, $X_{ql\_be\_s}$, $X_{qn\_bne\_s}$, and $X_{ql\_cne\_s}$, we incorporated corresponding variants with moderate (\lq\lq \_m''’) and weak (\lq\lq \_w'') effects. Here, the abbreviations \lq\lq bne'', \lq\lq be'', and \lq\lq cne'' indicate the type of categorical covariate involved in the interaction: \lq\lq bne'' refers to a binary covariate without a marginal effect, \lq\lq be'' to a binary covariate with a marginal effect, and \lq\lq cne'' to a categorical covariate with three levels and no marginal effect (see Figure~\ref{fig:sim_des_2_example_data}).
The 19 informative covariates in DGP~2 result from including one representative covariate for each considered effect type.

A detailed description of this DGP is provided in Section~C.2.2 of the supplementary material.

\subsubsection{Evaluation}

Two metrics were used to evaluate the results: AUC values and median ranks.

The AUC quantified how well each VIM distinguished informative from non-informative covariates. For each informative covariate type, we calculated an AUC value representing the probability that the VIM ranks a covariate of this type higher than a non-informative one. Higher AUC values therefore indicate a stronger ability of the VIM to separate informative from non-informative covariates. To compute these AUC values, we calculated an AUC per informative covariate type for each simulated dataset, using the VIM scores as the score variable. In this calculation, the first group comprised the representative(s) of the respective covariate type in the DGP, and the second group comprised the 50 non-informative covariates. Because the first group contained only two observations in DGP~1 and only one observation in DGP~2, the resulting AUC values naturally exhibited high variance. However, averaging these AUC values across all 1,000 simulated datasets (separately by sample size) greatly reduced this variance. To quantify the remaining uncertainty, 95\% confidence intervals were computed using the estimated standard error of the mean, relying on the asymptotic normality of the sample mean.

To assess how differently the VIMs rank covariates with different effect types, we also analyzed the ranks of the covariates among all available covariates. Specifically, for each simulated dataset, we determined the ranks of all covariates based on their VIM values. For each informative covariate type, we then computed the median rank across all datasets (stratified by sample size). To assess the variability of these ranks, we additionally calculated the first and third quartiles of the ranks across datasets. Lower ranks indicate higher importance as assigned by the respective VIM.

\subsection{Results}

In the following, we report the results in terms of both the AUC values and the median ranks. These two measures capture different aspects of performance but exhibited largely consistent patterns in our simulations: high AUC values were typically associated with low median ranks, and vice versa. Consequently, we focus primarily on the AUC results, referring to the rank-based results only where they offer additional insights.

A detailed description of the results is provided below, while Section~\ref{sec:sim_summary} presents a concise summary.

\subsubsection{DGP~1}

\begin{figure}[!ht]
\centering
\includegraphics[width=\linewidth]{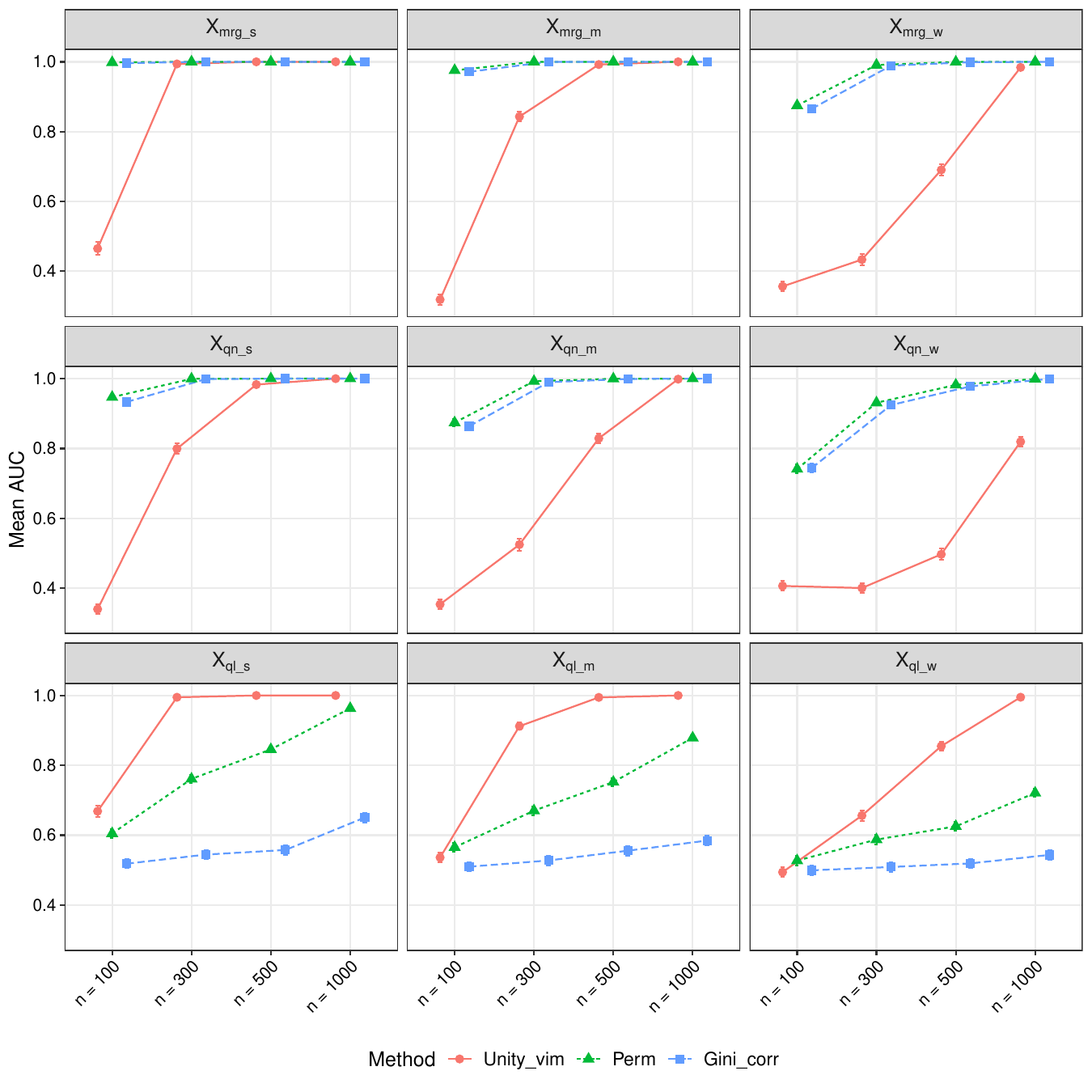}
\caption{Mean AUC values with 95\% confidence intervals per considered sample size and method for DGP~1}
\label{fig:sim_dgp1_auc}
\end{figure}

\begin{figure}[!ht]
\centering
\includegraphics[width=\linewidth]{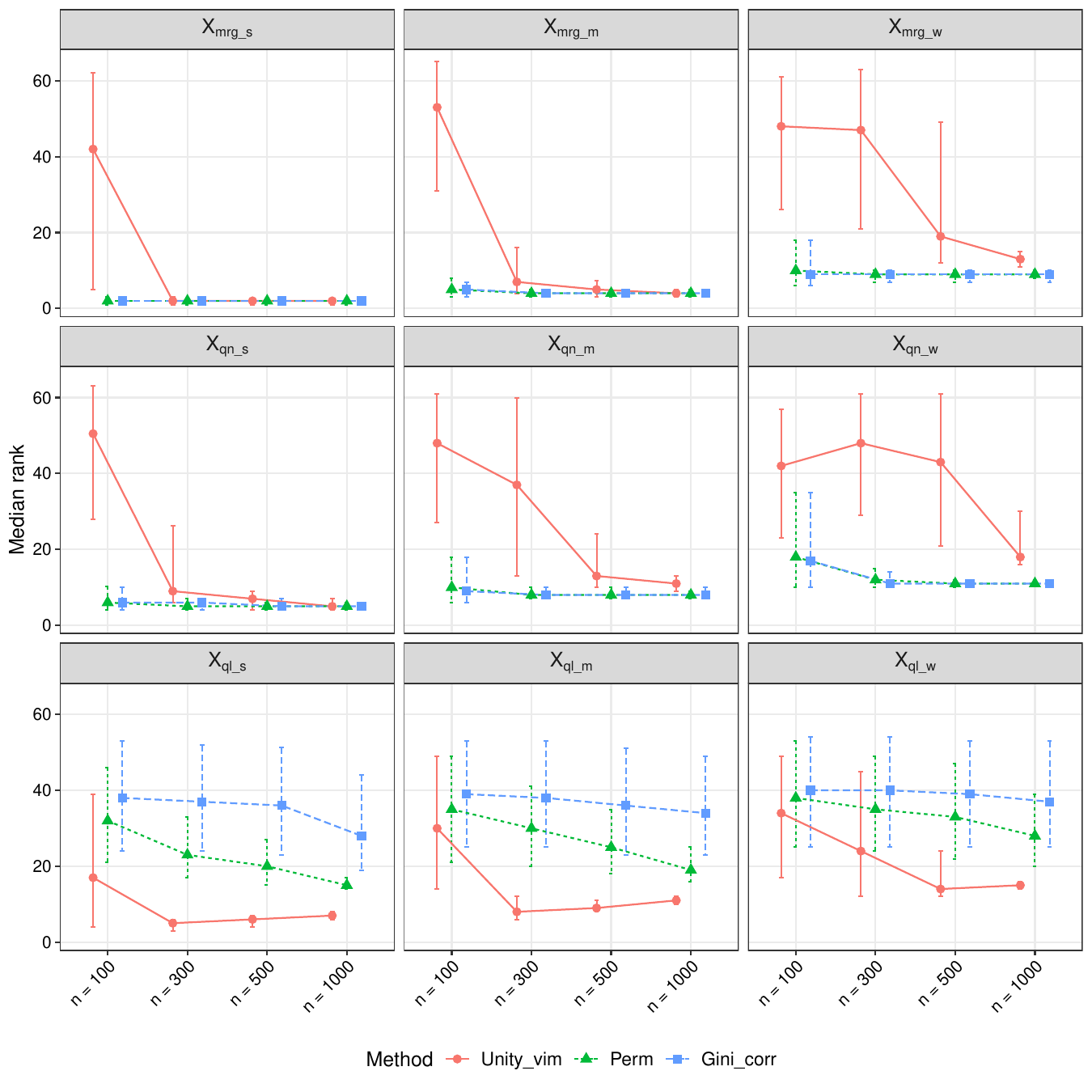}
\caption{Median ranks with 25\% and 75\% quartiles per considered sample size and method for DGP~1}
\label{fig:sim_dgp1_rank}
\end{figure}

Figure~\ref{fig:sim_dgp1_auc} shows the AUC values, and Figure~\ref{fig:sim_dgp1_rank} presents the corresponding ranks. The numerical values are reported in Tables~S6 and~S7 of the supplementary material.

A general observation is that, for the smallest sample size considered ($n = 100$), \rcode{Unity\_vim} did not rank any of the informative covariates higher than the non-informative ones. Most AUC values achieved with \rcode{Unity\_vim} were close to~0.5, with a maximum of~0.71. While the AUC values obtained with \rcode{Perm} and \rcode{Gini} for $n = 100$ were, almost without exception, higher than those of \rcode{Unity\_vim}, even these VIMs clearly separated informative from non-informative covariates only in the case of covariates with purely marginal and at the same time strong or moderate effects. The fact that \rcode{Unity\_vim} did not (reliably) rank any informative covariates above the non-informative ones for $n = 100$ suggests that small datasets may not be suitable for applying the unity VIM. The relatively high algorithmic complexity could lead to substantial variability in the results for small sample sizes; see Section~\ref{sec:discussion} for a detailed discussion. The subsequent descriptions of the results for DGP~1 therefore refer exclusively to settings with sample sizes greater than $n = 100$.

The covariates with purely marginal effects ($X_{mrg\_s}$, $X_{mrg\_m}$, and $X_{mrg\_w}$) were clearly distinguished from the non-informative covariates by both \rcode{Perm} and \rcode{Gini}. \rcode{Unity\_vim} also ranked these covariates reliably higher than the non-informative ones, with exceptions for the weak and moderate effects in smaller-sample size settings.

With regard to the covariates exhibiting quantitative interaction effects and marginal effects ($X_{qn\_s}$, $X_{qn\_m}$, and $X_{qn\_w}$), \rcode{Unity\_vim} performed comparably to or worse than the conventional VIMs, depending on the effect size and sample size. The conventional VIMs consistently achieved higher values for this covariate type than for the non-informative covariates, except for the weak effect at smaller sample sizes. In contrast, \rcode{Unity\_vim} reliably ranked this covariate type higher than the non-informative covariates only for the largest sample size and the moderate or strong effect. For the second-largest sample size, \rcode{Unity\_vim} provided a clear separation only in the case of the strong effect. In all remaining cases, \rcode{Unity\_vim} did not consistently distinguish this covariate type from the non-informative covariates. These findings are also largely reflected in the rank-based results (Figure~\ref{fig:sim_dgp1_rank}). Notably, the ranks of the weak-effect covariates are very high for \rcode{Unity\_vim}, confirming that it fails to separate weak quantitative interaction effects of this type from noise covariates. For the moderate effect and the second-largest sample size, the median rank of~13 achieved by \rcode{Unity\_vim} is lower than one might expect based on the relatively modest AUC value of~0.83; however, the variability of the corresponding ranks is considerable.

When detecting covariates with qualitative interaction effects and no marginal effects ($X_{ql\_s}$, $X_{ql\_m}$, and $X_{ql\_w}$), \rcode{Unity\_vim} consistently outperformed the conventional VIMs. For strong and moderate effects, \rcode{Unity\_vim} reliably ranked this covariate type above the non-informative covariates, with only a minor exception at $n = 300$ for moderate effects, where the AUC dropped slightly to~0.91. For the weak effect, \rcode{Unity\_vim} achieved clear separation from the non-informative covariates only for the largest sample size.

By contrast, \rcode{Perm} achieved a clear separation for this covariate type only for the largest sample size and only for the strong effect; for the moderate effect, separation was partial (AUC = 0.88). In all other cases, \rcode{Perm} did not allow for a clear distinction. For \rcode{Gini}, the AUC values for this covariate type were consistently close to~0.5, indicating that \rcode{Gini} failed to distinguish the covariate type from non-informative covariates. Section~\ref{sec:discussion} provides a possible explanation for the performance differences between \rcode{Perm} and \rcode{Gini} in detecting interaction effects without marginal effects.

Although \rcode{Perm} showed limited success in identifying this covariate type, its median ranks were never below~15, suggesting that such covariates would likely be overlooked when focusing on the top-ranked covariates. In contrast, the median ranks obtained with \rcode{Unity\_vim} for this covariate type were below~10 for both strong and moderate effects (with one exception).

\subsubsection{DGP~2}

\begin{figure}[!ht]
\centering
\includegraphics[width=\linewidth]{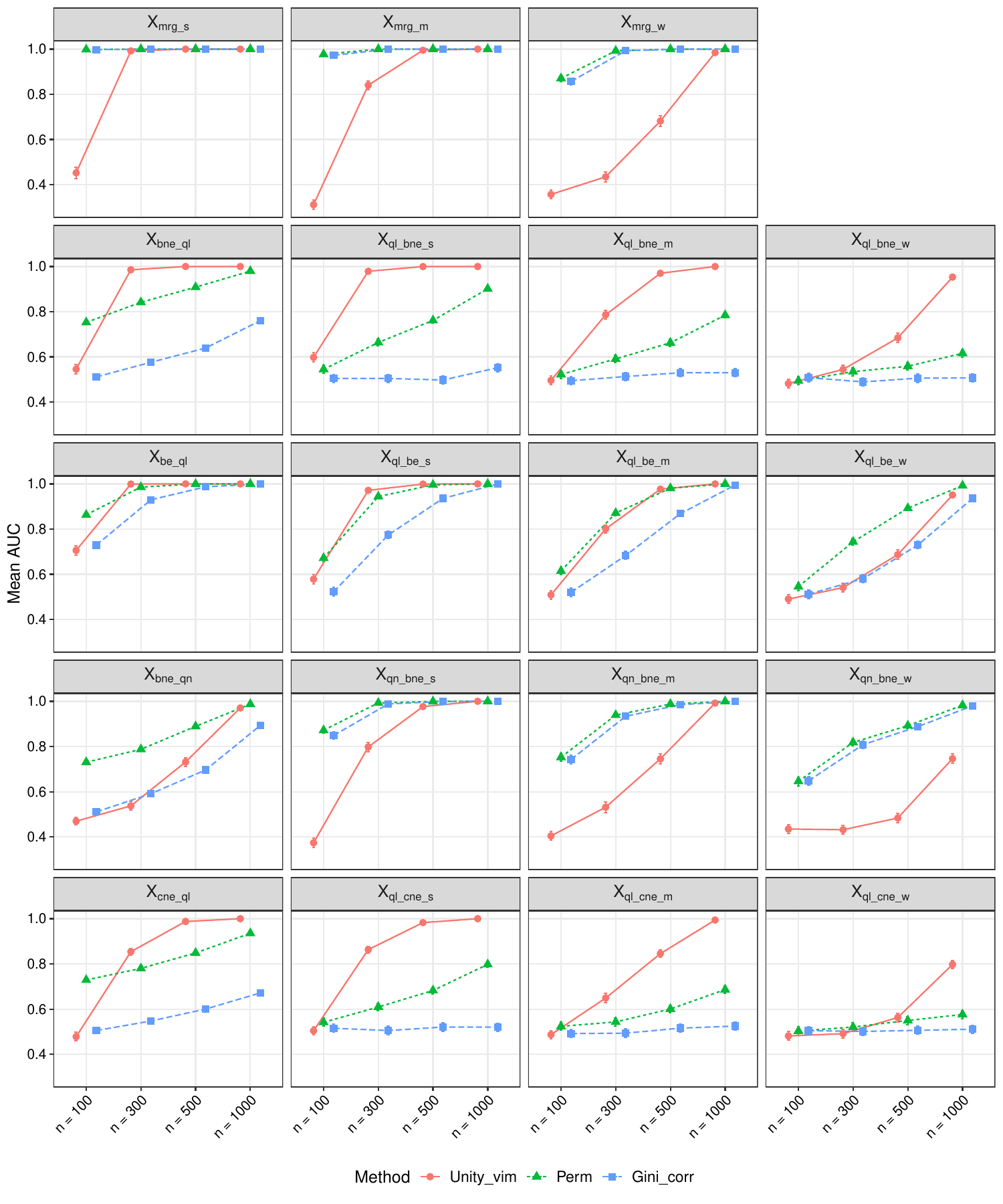}
\caption{Mean AUC values with 95\% confidence intervals per considered sample size and method for DGP~2}
\label{fig:sim_dgp2_auc}
\end{figure}

The results for DGP~2 are displayed in Figures~\ref{fig:sim_dgp2_auc} and \ref{fig:sim_dgp2_rank}. The underlying numerical results are provided in Tables~S8--S11 of the supplementary material.

\begin{figure}[!ht]
\centering
\includegraphics[width=\linewidth]{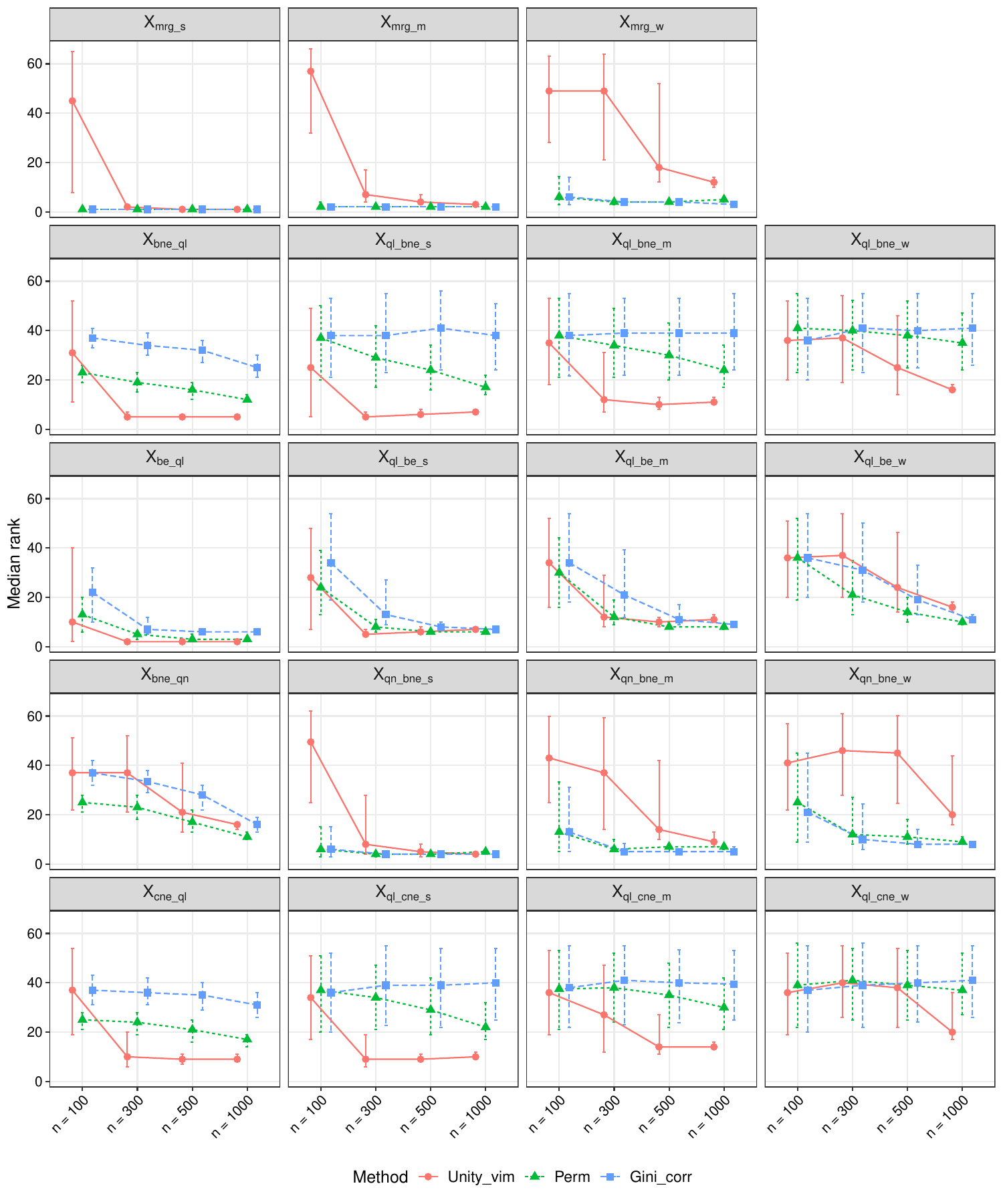}
\caption{Median ranks with 25\% and 75\% quartiles per considered sample size and method for DGP~2}
\label{fig:sim_dgp2_rank}
\end{figure}

For the smallest sample size ($n = 100$), the findings mirror those of DGP~1. In this setting, \rcode{Unity\_vim} did not separate any of the informative covariates from the non-informative ones. In contrast, the conventional VIMs consistently ranked the covariates with purely marginal effects higher than the non-informative covariates for the strong and moderate effects, though they did not allow for a clear separation in the remaining cases. As in the previous subsection, the following discussion refers exclusively to settings with sample sizes greater than $n = 100$.

The results for the covariates with purely marginal effects ($X_{mrg\_s}$, $X_{mrg\_m}$, and $X_{mrg\_w}$) closely resemble those from DGP~1. All VIMs reliably ranked these covariates higher than the non-informative ones, with the exception of \rcode{Unity\_vim} for the weak and moderate effects at smaller sample sizes.

To streamline the discussion, we omit the subscripts \lq\lq \_s'', \lq\lq \_m'', and \lq\lq \_w'' when referring to specific covariate types in the following; for example, $X_{ql\_bne}$ collectively denotes the covariate type represented by $X_{ql\_bne\_s}$, $X_{ql\_bne\_m}$, and $X_{ql\_bne\_w}$.

The binary covariate without a marginal effect, $X_{bne\_ql}$---which participates in a qualitative interaction effect without marginal effects together with the continuous covariate $X_{ql\_bne}$---was consistently ranked higher than the non-informative covariates by \rcode{Unity\_vim}. In contrast, \rcode{Perm} reliably distinguished this covariate from the non-informative ones only for the largest sample size, while \rcode{Gini} failed to do so for any of the sample sizes considered.

Similarly, \rcode{Unity\_vim} outperformed the conventional VIMs for the continuous covariate $X_{ql\_bne}$. The strong effect was reliably distinguished from the non-informative covariates by \rcode{Unity\_vim} across all sample sizes; the moderate effect was reliably separated at larger sample sizes; and even the weak effect was distinguished approximately reliably (AUC = 0.95) at the largest sample size. In contrast, \rcode{Perm} ranked this covariate type higher than the non-informative covariates only for the largest sample size and only in the case of the strong effect, and even then not consistently (AUC = 0.90). In all other cases, \rcode{Perm} did not rank this covariate type substantially higher than noise covariates. The AUC values obtained with \rcode{Gini} were consistently close to~0.5, indicating that this VIM failed to distinguish this covariate type from the non-informative covariates.

The qualitative interaction effect between the binary covariate $X_{be\_ql}$ and the continuous covariate $X_{ql\_be}$ resembles that between $X_{bne\_ql}$ and $X_{ql\_bne}$, with the difference that $X_{be\_ql}$ additionally has a moderate marginal effect. \rcode{Unity\_vim} reliably separated $X_{be\_ql}$ from the non-informative covariates, similar to its performance for $X_{bne\_ql}$. In contrast to $X_{bne\_ql}$, however, the conventional VIMs also achieved clear separation for $X_{be\_ql}$---likely due to the presence of the marginal effect. The only exception occurred for $n = 300$, where \rcode{Gini} ranked this covariate higher than the non-informative covariates not consistently, only very frequently (AUC = 0.93). The difference in performance between $X_{be\_ql}$ and $X_{bne\_ql}$ is also evident from the ranks (Figure~\ref{fig:sim_dgp2_rank}): whereas $X_{bne\_ql}$ received relatively high rank values for the conventional VIMs, $X_{be\_ql}$ consistently ranked among the top ten covariates for these VIMs. \rcode{Unity\_vim} ranked both $X_{bne\_ql}$ and $X_{be\_ql}$ within the top ten, assigning even higher ranks to $X_{be\_ql}$ than the conventional VIMs.

For the continuous covariate $X_{ql\_be}$, \rcode{Unity\_vim} performed similarly to \rcode{Perm}. For larger sample sizes, both methods reliably distinguished this covariate from the non-informative ones in the cases of the moderate or strong effects. In the case of the weak effect, \rcode{Unity\_vim} yielded considerably lower AUC values than \rcode{Perm}. However, for the weak effect, even \rcode{Perm} only achieved clear separation for the largest sample size (AUC = 0.99), where \rcode{Unity\_vim} also performed well (AUC = 0.95). \rcode{Gini} generally performed worse than both \rcode{Unity\_vim} and \rcode{Perm}, achieving consistent separation only for the largest sample size and only for the strong and moderate effects.

The binary covariate $X_{bne\_qn}$, involved in a quantitative interaction with the continuous covariate $X_{qn\_bne}$, was consistently ranked higher than the non-informative covariates only by \rcode{Perm}, and nearly consistently by \rcode{Unity\_vim} (AUC = 0.97), both for the largest sample size. For smaller sample sizes, no clear separation was observed, although \rcode{Perm} achieved substantially higher AUC values than \rcode{Unity\_vim} and \rcode{Gini}. For $X_{qn\_bne}$, both \rcode{Perm} and \rcode{Gini} reliably distinguished this covariate from the non-informative ones for the strong effect across the considered sample sizes, for the moderate effect at larger sample sizes, and for the weak effect only at the largest sample size. The same general pattern was observed for \rcode{Unity\_vim}, although less pronounced: $X_{qn\_bne}$ was consistently ranked higher only for the strong effect at larger sample sizes, for the moderate effect only at the largest sample size, and not at all for the weak effect.

The categorical covariate with three levels, $X_{cne\_ql}$---which has no marginal effect but participates in a qualitative interaction with the continuous covariate $X_{ql\_cne}$, also without a marginal effect---was clearly separated from the non-informative covariates only by \rcode{Unity\_vim}, and only for larger sample sizes. The conventional VIMs failed to consistently distinguish this covariate from the non-informative ones for any sample size, although \rcode{Perm} performed somewhat better, ranking this covariate higher very frequently (but not consistently) at the largest sample size (AUC = 0.94).

For the continuous covariate $X_{ql\_cne}$, \rcode{Unity\_vim} again outperformed both conventional VIMs. \rcode{Unity\_vim} reliably ranked this covariate higher than the non-informative ones for the largest sample size in the case of the strong and moderate effects, and for the second-largest sample size in the case of the strong effect. As shown in Figure~\ref{fig:sim_dgp2_rank}, both $X_{ql\_cne}$ and $X_{cne\_ql}$ tended to receive somewhat lower ranks compared to the other informative covariate types. The conventional VIMs, by contrast, failed to reliably separate $X_{ql\_cne}$ from the non-informative covariates for any effect size or sample size, with \rcode{Gini} in particular producing AUC values consistently close to~0.5.

\subsection{Summary and Conclusions}
\label{sec:sim_summary}

In contrast to the conventional VIMs, the unity VIM reliably distinguished covariates involved in interaction effects without marginal effects from non-informative covariates. The corrected Gini importance failed to identify such covariates for any of the effect sizes or sample sizes considered, while the permutation importance frequently---but not consistently---ranked them higher than the non-informative covariates in the presence of strong effects and large sample sizes.

Covariates without marginal effects that interacted with covariates possessing marginal effects were clearly identified as influential by all VIMs.

For smaller sample sizes and weaker effect strengths, the unity VIM was less reliable than the conventional VIMs in distinguishing covariates with marginal effects from non-informative covariates. A plausible explanation is the higher variability of the unity VIM, which likely stems from the greater algorithmic complexity of UFOs. This variability is also evident in the rank-based results (Figures~\ref{fig:sim_dgp1_rank} and~\ref{fig:sim_dgp2_rank}), where the interquartile ranges for the unity VIM are generally larger than those for the conventional VIMs across many covariates.

For the smallest sample size considered ($n = 100$), the unity VIM did not rank any of the informative covariate types higher than the non-informative covariates. This indicates that unity VIM results for very small datasets should be interpreted with caution.

\section{Empirical Evaluation of the CRTR Algorithm}
\label{sec:crtr_analysis}

In this section, we first apply the CRTR algorithm to two simulated datasets of size $n = 500$, generated using the two DGPs described in Section~\ref{sec:sim}, to examine the extent to which the association structures identified by the algorithm align with the true underlying structures. We then apply the algorithm to the well-known wine dataset ($n = 178$, $p = 13$) for illustrative purposes. In all analyses, we use the default hyperparameter values of the UFO framework described in Section~\ref{sec:ufo_framework} and in Sections~A.3 and~A.4 of the supplementary material.

\subsection{Evaluation on Simulated Data}
\label{sec:crtr_sim}

For reasons of space, we only present the CRTRs for the five covariates with the highest unity VIM values in the main paper. In Section~D.1 of the supplementary material, we show and describe the 20 highest unity and permutation VIM values for both simulated datasets, along with additional descriptions.

\begin{figure}[!ht]
\centering
\includegraphics[width=0.95\linewidth]{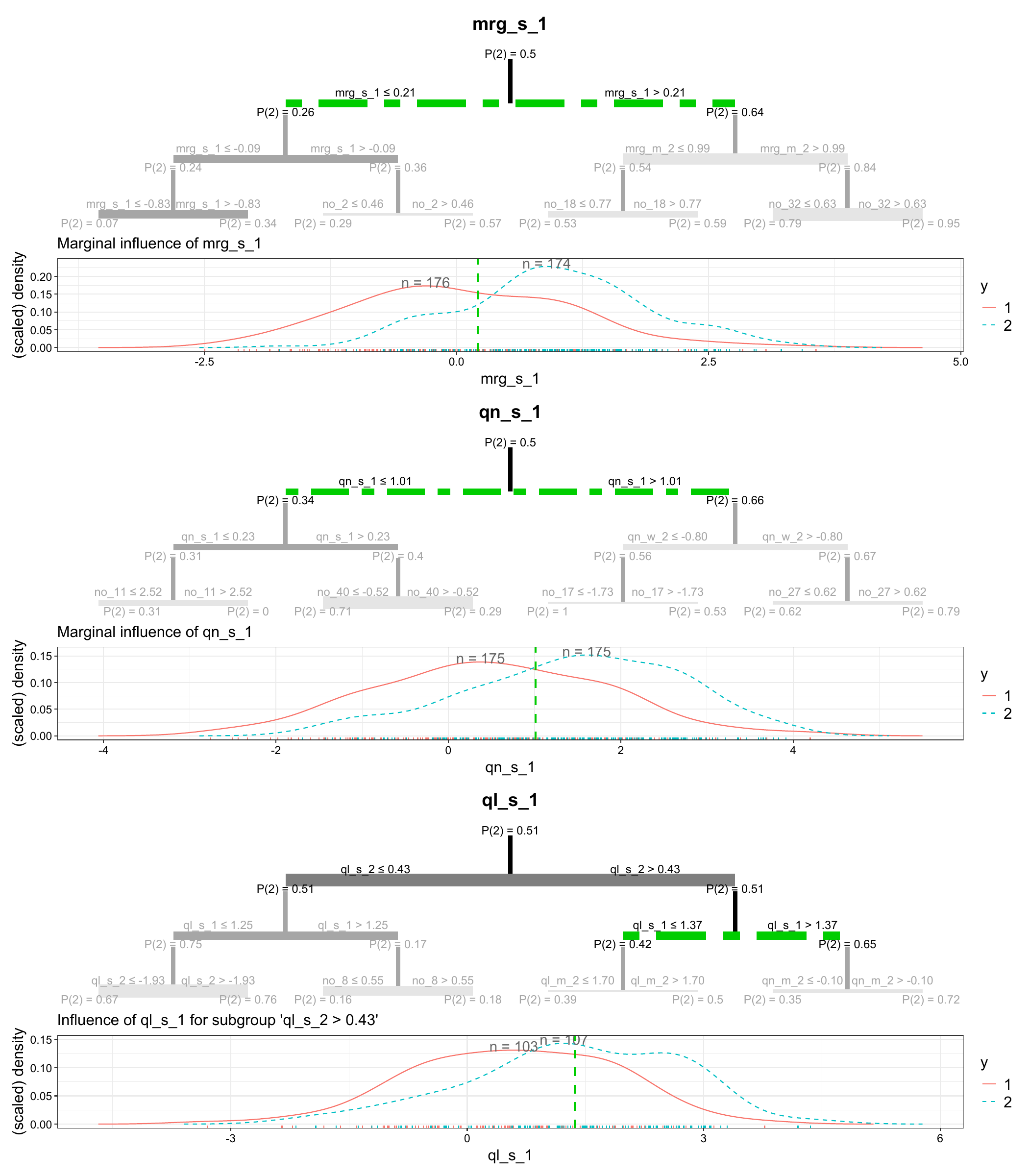}
\caption{Visualization of the CRTRs for the three covariates with the highest unity VIM values in a dataset generated from DGP~1 ($n=500$). The thickness of the horizontal lines reflects the covariate scores, and dashed lines indicate top-scoring splits. Gray-shaded areas mark tree regions that do not belong to top-scoring splits or their ancestors. The kernel density estimates of the class-specific covariate distributions in the top-scoring nodes were weighted by the class proportions. The class frequency distributions at the nodes and the kernel density estimates were computed using the in-bag data. $\text{P}(2)$ denotes the in-bag proportion of observations belonging to class~2 at the respective node. Covariate labels are abbreviated (e.g., \rcode{mrg\_s\_1} denotes $X_{mrg\_s\_1}$ and \rcode{no\_*} denotes non-informative covariates)}
\label{fig:crtrs_dgp1_1}
\end{figure}

\begin{figure}[!ht]
\centering
\includegraphics[width=0.95\linewidth]{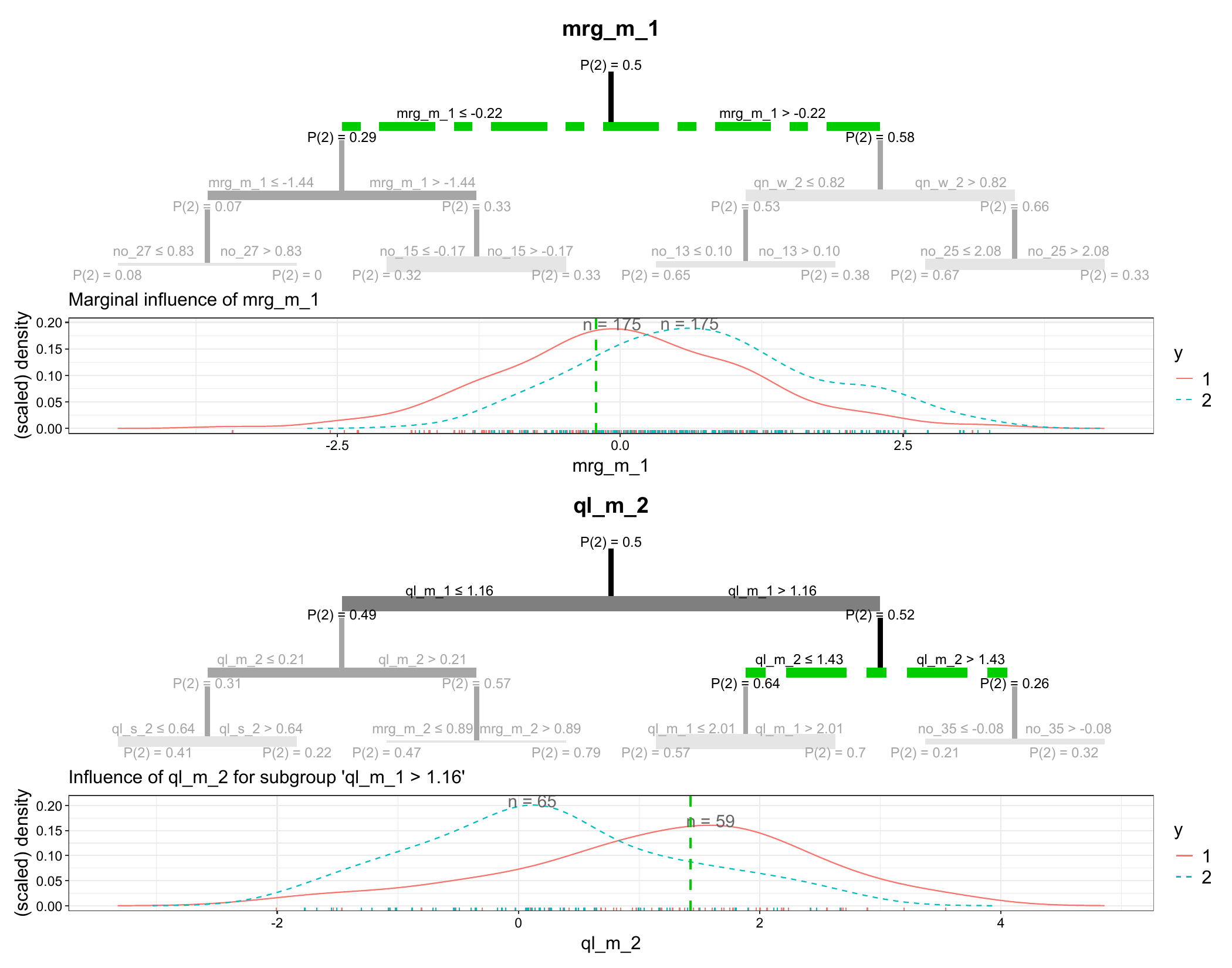}
\caption{Visualization of the CRTRs for the fourth and fifth highest-ranked covariates according to the unity VIM in a dataset generated from DGP~1 ($n=500$). The thickness of the horizontal lines reflects the covariate scores, dashed lines indicate top-scoring splits, and gray-shaded areas mark tree regions outside these splits and their ancestors. The class frequency distributions at the nodes and the weighted kernel density estimates of the class-specific covariate distributions were computed using the in-bag data. $\text{P}(2)$ denotes the in-bag proportion of observations belonging to class~2 at the respective node. Covariate labels are abbreviated (e.g., \rcode{mrg\_m\_1} denotes $X_{mrg\_m\_1}$ and \rcode{no\_*} denotes non-informative covariates)}
\label{fig:crtrs_dgp1_2}
\end{figure}

Figures~\ref{fig:crtrs_dgp1_1} and~\ref{fig:crtrs_dgp1_2} show the CRTRs for the five covariates with the highest unity VIM values from the dataset generated with DGP~1, as well as kernel density estimates of the class-specific covariate distributions in the nodes corresponding to the top-scoring splits of the respective CRTRs.\footnote{As described in Section~\ref{sec:sim_design}, DGP~1 contains two covariates per covariate type and effect size. To avoid duplication, in cases where both members of a pair were among the five highest-ranked covariates, the covariate with the lower unity VIM value was excluded. Thus, each CRTR shown represents a different combination of covariate type and effect size.}

Before discussing the substantive interpretation, we first describe the visualization of CRTRs used in this section, which is also implemented in a slightly modified form in the R package \rcode{unityForest}. In each visualization, the relative class frequency distributions of the in-bag observations at the nodes are displayed. For binary outcomes, as in the present example, it is sufficient to show the frequencies for one of the two classes.

The thickness of the horizontal lines corresponds to the covariate scores described in the detailed description of the algorithm in Section~\ref{sec:crtrs}. These scores reflect how much more frequently the respective split covariates occur in the best tree roots of the covariate under consideration compared to their frequency across all tree roots in the forest. A particularly thick horizontal line for a split covariate therefore indicates that this covariate appears especially often in tree roots where the covariate for which the CRTR was obtained has a strong effect. This suggests that the covariate is likely interacting with the covariate of interest.

Top-scoring splits are indicated by dashed horizontal lines. For interpretability it is helpful to shade out the areas of the CRTR that do not belong to the nodes with top-scoring splits or their ancestors, as in the figures presented here. Two considerations motivate this approach. First, the objective of the CRTRs is to depict the conditions under which the respective covariates exert their strongest effects---conditions that are determined by the ancestors of the nodes containing the top-scoring splits. Second, the regions of the CRTRs not belonging to the nodes with top-scoring splits or their ancestors are of limited interpretive value. Each CRTR is derived from tree roots in which the respective covariate exhibits a particularly strong effect. Consequently, the association patterns that frequently occur within these tree roots are specific to that covariate. In contrast, the shaded regions of the CRTRs reflect arbitrary aspects of the overall association structure in the data. Moreover, these regions may also include splits on non-informative covariates, since each tree root is constructed from a small random subset of all available covariates---many of which may be non-informative.

At the same time, splits on the covariate under consideration may also occur within the shaded regions and can be relevant for interpretation. However, these splits do not belong to the top-scoring splits defining the CRTR and therefore do not (generally) represent the conditions under which the covariate exhibits its strongest effects.

Finally, the kernel density estimates of the class-specific covariate distributions in the nodes with the top-scoring splits help to assess the direction and strength of the covariate effects. For better interpretability, the densities are scaled by the proportions of observations from the respective classes, allowing them to be interpreted as local class-specific densities relative to one another.

The CRTRs shown in Figures~\ref{fig:crtrs_dgp1_1} and~\ref{fig:crtrs_dgp1_2} reflect the true types of effects of the corresponding covariates well. The quantitative interaction effect between $X_{qn\_s\_1}$ and $X_{qn\_s\_2}$, both of which also exhibit marginal effects, is not reflected by the CRTR of $X_{qn\_s\_1}$. However, the purpose of the unity VIM and the CRTRs is not to detect every possible kind of interaction effect but to identify the conditions under which the corresponding covariates exert their strongest influence. $X_{qn\_s\_1}$ not only has a strong effect for observations with small $X_{qn\_s\_2}$ values due to the interaction but also a marginal effect. Combined with the fact that the top-scoring splits are weighted by node size, this likely led to the top-scoring split in $X_{qn\_s\_1}$ appearing as the first split in the CRTR---thus reflecting its marginal rather than interaction-based influence.

The CRTRs of $X_{mrg\_s\_1}$ and $X_{mrg\_m\_1}$, which have purely marginal effects, correctly represent their  influence types. Likewise, the CRTRs of $X_{ql\_s\_1}$ and $X_{ql\_m\_2}$ correctly capture their qualitative interaction effects. These covariates have no marginal effects but interact with $X_{ql\_s\_2}$ and $X_{ql\_m\_1}$, respectively, such that the direction of their effects reverses depending on whether $X_{ql\_s\_2}$ or $X_{ql\_m\_1}$ take small or large values. This structure is clearly visible in the corresponding CRTRs: in the CRTR of $X_{ql\_s\_1}$, a split first occurs in $X_{ql\_s\_2}$, followed by splits in $X_{ql\_s\_1}$ within both child nodes of the root node. The same pattern appears in the CRTR of $X_{ql\_m\_2}$. Figures~S6–S14 in the supplementary material display the CRTRs and top-scoring split visualizations for all informative covariates in the DGP~1 dataset. These results are consistent with those shown in the main paper and further demonstrate that even covariates with weak effects were correctly characterized by their CRTRs.

\begin{figure}[!ht]
\centering
\includegraphics[width=0.95\linewidth]{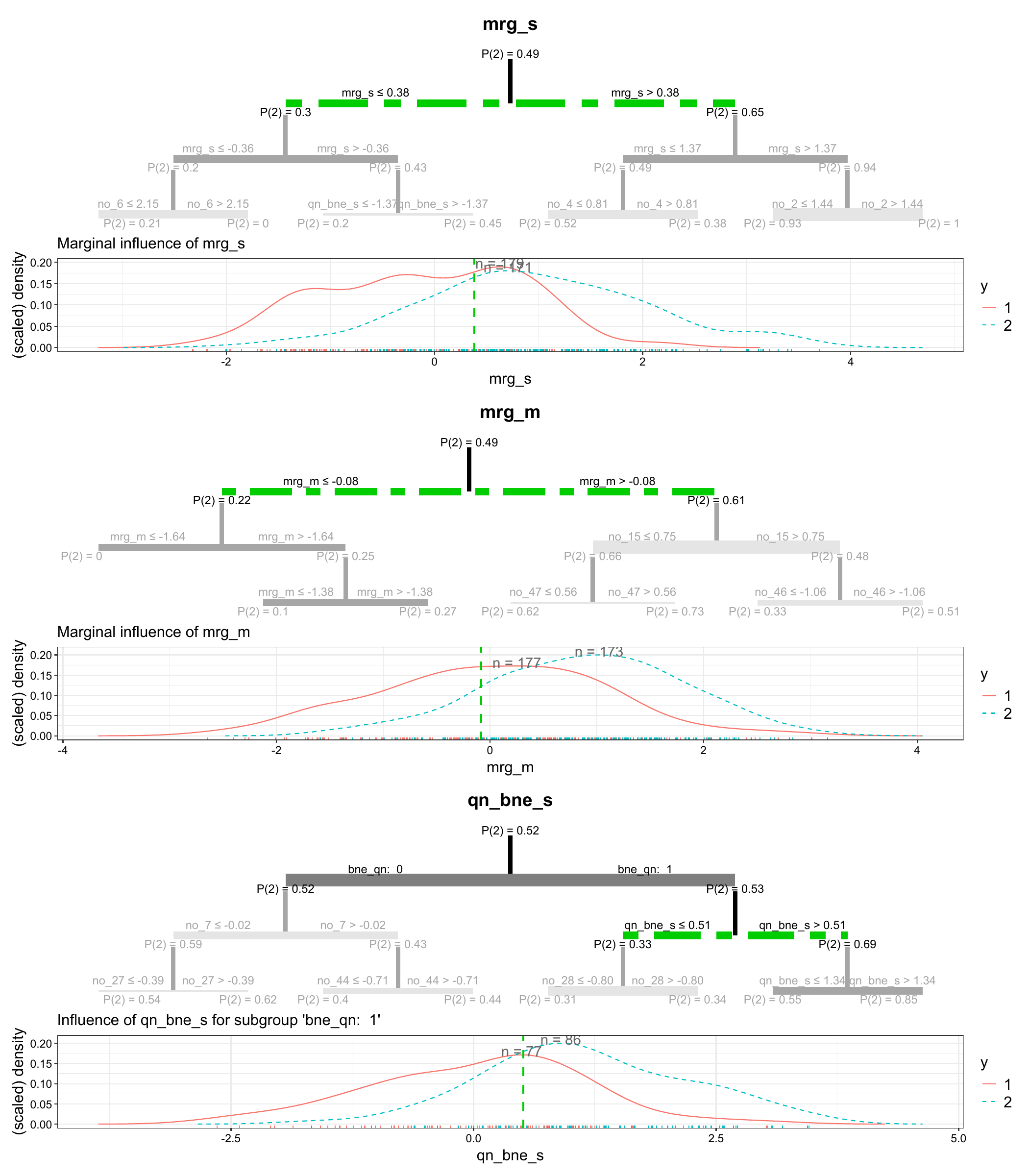}
\caption{Visualization of the CRTRs for the three covariates with the highest unity VIM values in a dataset generated from DGP~2 ($n=500$). The thickness of the horizontal lines reflects the covariate scores, dashed lines indicate top-scoring splits, and gray-shaded areas mark tree regions outside these splits and their ancestors. The class frequency distributions at the nodes and the weighted kernel density estimates of the class-specific covariate distributions were computed using the in-bag data. $\text{P}(2)$ denotes the in-bag proportion of observations belonging to class~2 at the respective node. Covariate labels are abbreviated (e.g., \rcode{mrg\_s} denotes $X_{mrg\_s}$ and \rcode{no\_*} denotes non-informative covariates)}
\label{fig:crtrs_dgp2_1}
\end{figure}

\begin{figure}[!ht]
\centering
\includegraphics[width=0.95\linewidth]{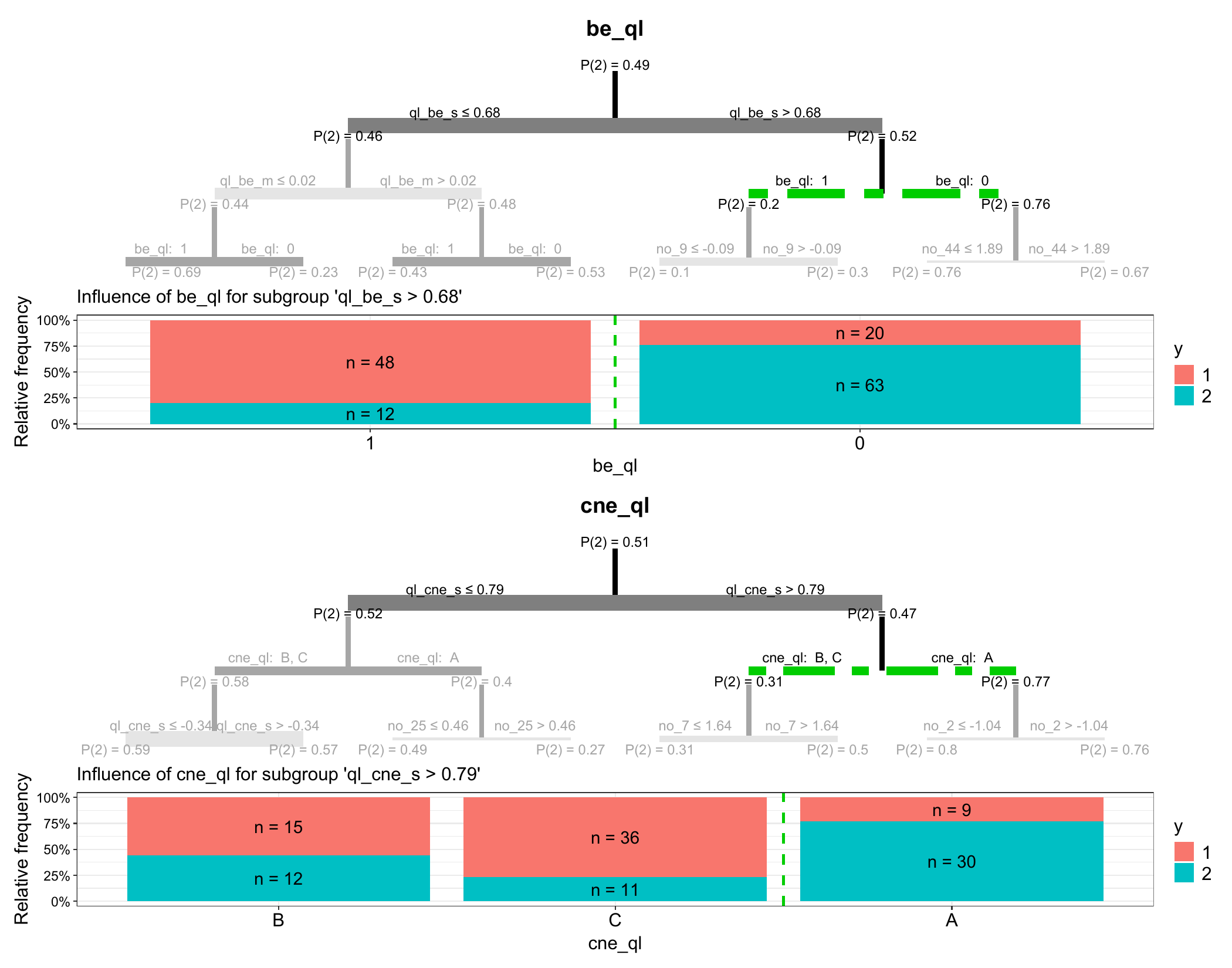}
\caption{Visualization of the CRTRs for the fourth and fifth highest-ranked covariates according to the unity VIM in a dataset generated from DGP~2 ($n=500$). The thickness of the horizontal lines reflects the covariate scores, dashed lines indicate top-scoring splits, and gray-shaded areas mark tree regions outside these splits and their ancestors. The class frequency distributions at the nodes and the bar plots were computed using the in-bag data. $\text{P}(2)$ denotes the in-bag proportion of observations belonging to class~2 at the respective node. Covariate labels are abbreviated (e.g., \rcode{be\_ql} denotes $X_{be\_ql}$ and \rcode{no\_*} denotes non-informative covariates)}
\label{fig:crtrs_dgp2_2}
\end{figure}

For the dataset generated under DGP~2, the CRTRs for the five covariates ranked highest by the unity VIM are shown in Figures \ref{fig:crtrs_dgp2_1} and \ref{fig:crtrs_dgp2_2}. Here, too, the CRTRs reflect the true effects of the corresponding covariates. In the CRTRs of the covariates with purely marginal effects, $X_{mrg\_s}$ and $X_{mrg\_m}$, the top-scoring splits correctly appear as the first splits.

In the CRTR of $X_{qn\_bne\_s}$, which exhibits a quantitative interaction effect with the binary covariate $X_{bne\_s}$, the top-scoring split---unlike for the quantitatively interacting covariate $X_{qn\_s\_1}$ in DGP~1---is not the first split. Instead, the first split occurs in $X_{bne\_s}$, followed by the top-scoring split in $X_{qn\_bne\_s}$ within the child node where $X_{bne\_s}=1$. This structure is consistent with the DGP. The fact that, unlike in the case of the CRTR of $X_{qn\_s\_1}$ in DGP~1 (Figure~\ref{fig:crtrs_dgp1_1}), the top-scoring split is not the first one here can be attributed to differences in the respective quantitative interaction effects between the two DGPs. The conditional effect of $X_{qn\_bne\_s}$ given $X_{bne\_s}=1$ (DGP~2) is stronger than the conditional effect of $X_{qn\_s\_1}$ given small $X_{qn\_s\_2}$ values (DGP~1): when $X_{bne\_s}=1$, one class dominates for small values of $X_{qn\_bne\_s}$ and the other for large values, whereas for small $X_{qn\_s\_2}$ values, both classes are similarly represented for small $X_{qn\_s\_1}$ values, and only for large $X_{qn\_s\_1}$ does one class begin to dominate.

The fourth- and fifth-highest-ranked covariates, $X_{be\_ql}$ and $X_{cne\_ql}$, are categorical and exhibit qualitative interaction effects with $X_{ql\_be\_s}$ and $X_{ql\_cne\_s}$, respectively. This is also reflected in their CRTRs: an initial split occurs in $X_{ql\_be\_s}$ and $X_{ql\_cne\_s}$, followed by top-scoring splits in $X_{be\_ql}$ and $X_{cne\_ql}$, respectively. Although $X_{be\_ql}$ also has a marginal effect, its influence is considerably stronger among observations with high $X_{ql\_be\_s}$ values. For these categorical covariates, stacked bar charts rather than kernel density estimates are used to illustrate the relationship between the categorical covariates and the binary outcome.

Figures~S15–S23 in the supplementary material present the corresponding results for all informative covariates, analogous to the DGP ~1 analysis. These results confirm the patterns shown in the main paper: the CRTRs correctly reproduced the true effect types of the covariates. The only exception was the CRTR of $X_{ql\_cne\_w}$, where the weak interaction effect with the categorical covariate $X_{cne\_ql}$ was not captured.

\subsection{Application to the Wine Dataset}

The well-known wine dataset comprises the results of chemical analyses of 178 wines produced in the same region of Italy from three grape varieties (Barolo, Grignolino, and Barbera). The dataset was originally introduced by \citet{Forina:1984} and later described in detail by \citet{Forina:1986}. For each sample, 13 continuous chemical constituents were measured, which serve as covariates for distinguishing between the grape varieties. For our analysis, we used a version of this dataset with a binary outcome that differentiates between Grignolino (\rcode{"G"}) and the two other varieties (Barolo and Barbera). This version is available on OpenML under data ID 973. The covariates include alcohol (\rcode{Alc}), malic acid (\rcode{Mal}), ash (\rcode{Ash}), alkalinity of ash (\rcode{AlcAsh}), magnesium (\rcode{Mg}), total phenols (\rcode{TP}), flavonoids (\rcode{Fla}), nonflavonoid phenols (\rcode{NFP}), proanthocyanins (\rcode{ProAn}), color intensity (\rcode{Col}), hue (\rcode{Hue}), OD280/OD315 of diluted wines (wine absorbance index) (\rcode{WAI}), and proline (\rcode{Prol}).

As in the analyses of the simulated data, we focus on the CRTRs for the five covariates with the highest unity VIM values. As mentioned in Section~\ref{sec:crtrs}, CRTRs are generally only informative for covariates with sufficiently high unity VIM values. The unity and permutation VIM values obtained are displayed in Figure~S24 in the supplementary material. The rankings of the covariates based on these two VIMs are broadly similar, although some differences can be observed that are also reflected by the CRTRs presented below. The Spearman correlation between the unity and permutation VIM values is 0.85.

\begin{figure}[!ht]
\centering
\includegraphics[width=0.95\linewidth]{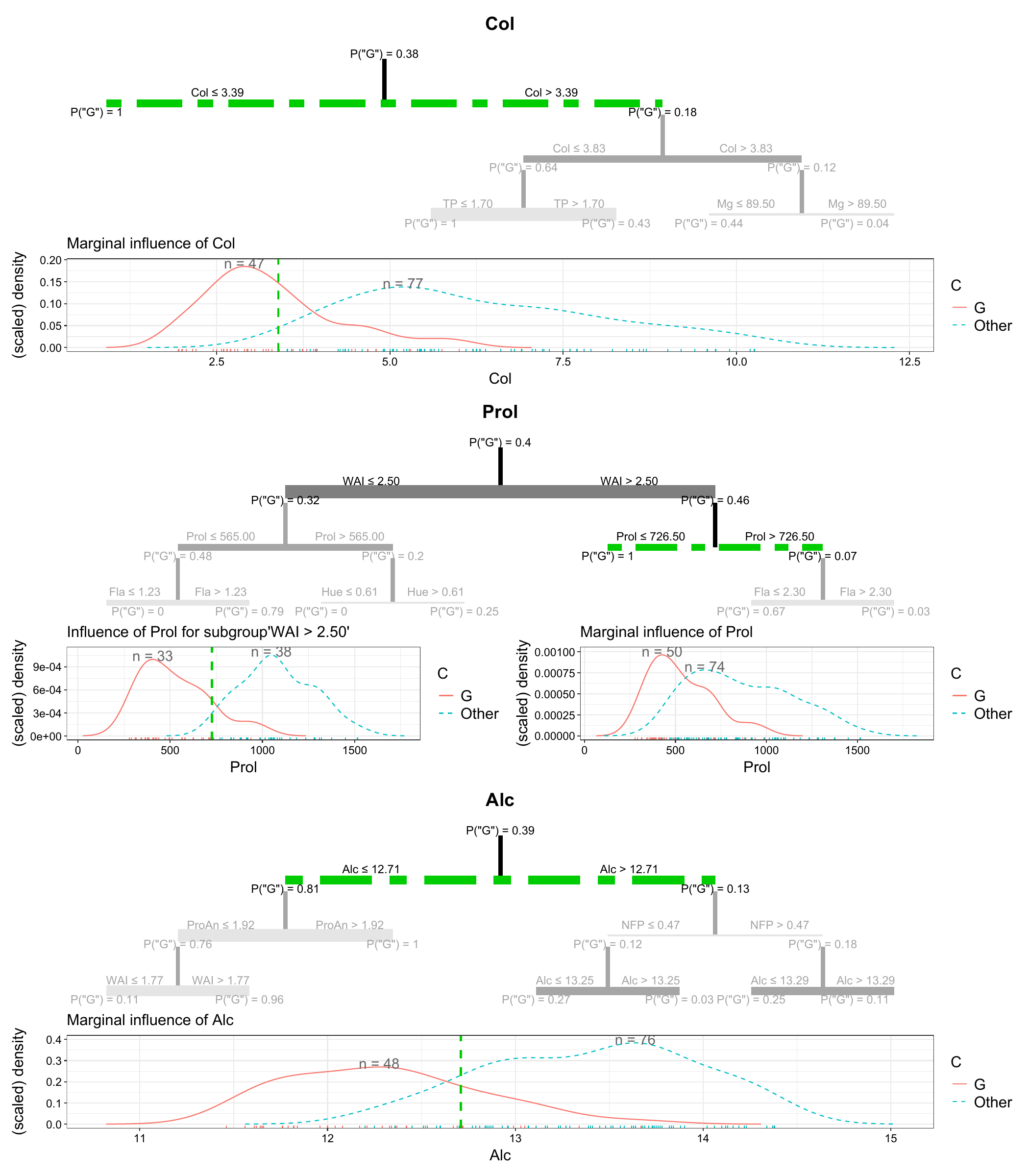}
\caption{Visualization of the CRTRs for the three covariates with the highest unity VIM values in the wine dataset ($n=178$). The thickness of the horizontal lines reflects the covariate scores, dashed lines indicate top-scoring splits, and gray-shaded areas mark tree regions outside these splits and their ancestors. For CRTRs where the first split is not a top-scoring split, kernel density estimates are shown both for all in-bag observations and for those in the top-scoring nodes. The class frequency distributions at the nodes and the weighted kernel density estimates of the class-specific covariate distributions were computed using the in-bag data. P(\lq\lq G'') denotes the in-bag proportion of observations belonging to Grignolino at the respective node}
\label{fig:crtrs_wine_1}
\end{figure}

\begin{figure}[!ht]
\centering
\includegraphics[width=0.95\linewidth]{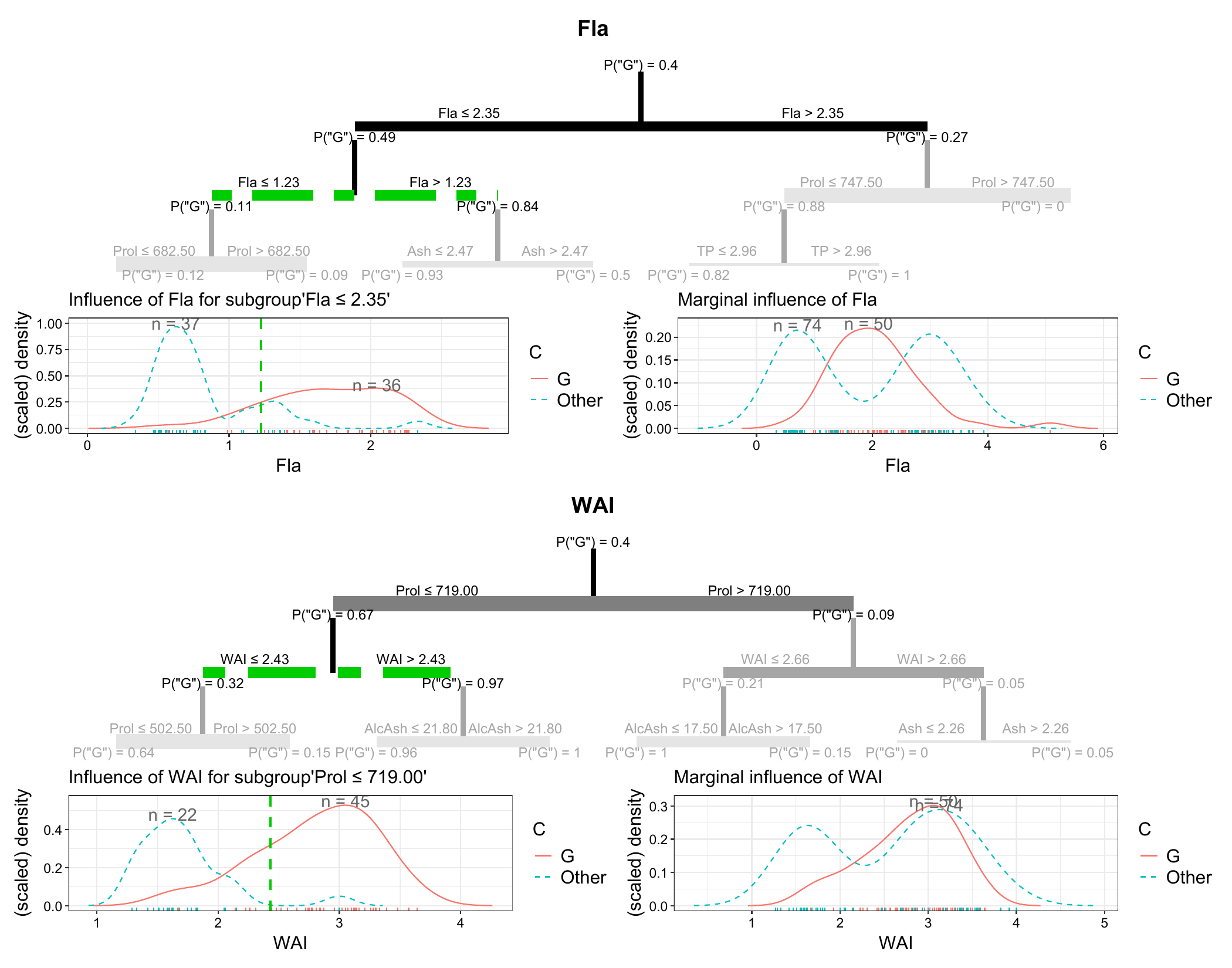}
\caption{Visualization of the CRTRs for the fourth and fifth highest-ranked covariates according to the unity VIM in the wine dataset ($n=178$). The thickness of the horizontal lines reflects the covariate scores, dashed lines indicate top-scoring splits, and gray-shaded areas mark tree regions outside these splits and their ancestors. For CRTRs where the first split is not a top-scoring split, kernel density estimates are shown both for all in-bag observations and for those in the top-scoring nodes. The class frequency distributions at the nodes and the weighted kernel density estimates of the class-specific covariate distributions were computed using the in-bag data. P(\lq\lq G'') denotes the in-bag proportion of observations belonging to Grignolino at the respective node}
\label{fig:crtrs_wine_2}
\end{figure}

The CRTRs for the five covariates ranked highest by the unity VIM are shown in Figures~\ref{fig:crtrs_wine_1} and~\ref{fig:crtrs_wine_2}. For the covariates color intensity and alcohol, the top-scoring splits occur as the first splits in their respective CRTRs, consistent with the strong marginal effects evident from the corresponding kernel density estimates.

In contrast, the CRTRs of proline and wine absorbance index exhibit different structures: in each case, the first split occurs in the other of the two covariates, followed by a split in the covariate of interest within both resulting child nodes. The kernel density estimates confirm that the conditional influences of these covariates are much stronger than their marginal effects. In particular, wine absorbance index shows only a weak marginal effect. The CRTR of this covariate also indicates a qualitative interaction between wine absorbance index and proline: while wine absorbance index appears to increase the probability of Grignolino for small proline values, it appears to have the opposite effect for large proline values. The observation that proline exhibits a stronger conditional than marginal effect is also reflected in the VIMs. While its unity VIM value is only slightly lower than that of color intensity, proline ranks third according to the permutation VIM, with a VIM value less than half that of color intensity.

The CRTR of flavonoids suggests a nonlinear association with the probability of Grignolino. Here, the first split occurs in flavonoids, and a second split---also in flavonoids---appears in the left child node of the root node, with only this second split being top-scoring. The frequency of Grignolino is low for both small and large flavonoid values but high for intermediate values. This pattern is evident both in the CRTR and in the estimated marginal effects from the kernel density plots.

\subsection{Conclusions}

The CRTRs obtained for the simulated datasets reflected the true effects of the informative covariates well throughout. The CRTR analysis of the wine dataset provided interesting insights into the effects of the covariates---insights that would likely remain hidden in conventional analyses. In particular, CRTRs offer a valuable means for interpreting VIM results in greater depth, as they capture both marginal and conditional covariate effects.

\clearpage
\section{Discussion}
\label{sec:discussion}

\myheading{Motivation and Central Idea of CRTRs}
Conventional parametric methods typically model covariate effects in a simplified, often linear manner and only capture interactions if these are explicitly specified. In practice, however, covariate effects may be more complex. The UFO framework enables the selection of CRTRs that identify the conditions under which covariates exert their strongest effects---potentially conditional on specific ranges of values of other covariates---and measure these effects accordingly. In our application, the CRTRs allowed us to illustrate the presence of such more complex effect patterns, as some of the top-scoring splits occurred below the root level, pointing to strong effects that arise through interactions rather than marginally.

\myheading{More Complex Covariate Effects: Challenges and Possible Extension of the CRTR Algorithm}
Despite this capability, real-world association structures can be more complex. For instance, a single covariate may interact with multiple other covariates, giving rise to several distinct conditions under which it exerts strong effects. In its current form, however, the UFO framework is designed to identify only the single condition under which each covariate shows its strongest effect. This alone already allows for the detection of effects that would remain hidden to methods focusing solely on the overall or marginal influence of covariates. Moreover, complex effects are inherently more challenging to interpret and communicate. A further practical limitation is that highly complex effect structures are difficult to identify: the statistical power to detect them is lower, and there is an increased risk that identified patterns represent non-generalizable artifacts in the observed data rather than true associations. This risk is comparatively low for CRTRs, as their objective is limited to identifying the strongest effects of each covariate.

Nevertheless, in some exploratory analyses, it may be of interest to obtain the most comprehensive possible picture of the covariate influences. One possible extension could involve selecting several heterogeneous CRTRs per covariate that capture distinct effect types. To achieve this, the respective set of best tree roots (see Section~\ref{sec:crtrs}) could be used to compute a distance matrix between them based on the measure proposed by \citet{Laabs:2024}. Hierarchical clustering could then be applied to this matrix, following the approach of \citet{Szepannek:2024}. The optimal number of clusters could be determined using the gap statistic \citep{Tibshirani:2001}, and from each resulting cluster, the CRTR with the smallest average distance to all other tree roots within that cluster could be selected as the representative structure.

\myheading{Relationship of Tree Roots to Optimal Trees and Their Properties}
As described in Section~\ref{sec:tree_constr}, the tree roots in UFOs are conceptually related to optimal trees. Since, in the construction of optimal trees, all split covariates and corresponding split points are selected simultaneously to achieve maximum leaf purity, such trees might be expected to have a tendency toward overfitting---potentially compromising the interpretability of the CRTRs.

However, the analyses by \citet{Bertsimas:2017}, who proposed an algorithm for constructing optimal classification trees, contradict this notion. In their results, optimal trees did not exhibit greater overfitting than CART trees. Moreover, overfitting did not increase with the number of covariates, and the sample sizes were small in all analyses. In a comparative study by \citet{Dunn:2021}, VIMs derived from optimal trees were less likely to classify non-informative covariates as informative than those based on CART trees. However, \citet{Dunn:2021} did not specify the exact type of VIMs used in their analyses. Their results also suggest that, unlike CART trees, optimal trees do not systematically favor covariates with many potential split points over those with fewer.

\myheading{Potential Use of Tree Roots in Boosting}
The principle used in UFOs to construct tree roots---selecting the best trees from a large number of randomly generated candidates---could also be promising for use in gradient boosting with trees. Because the trees used in gradient boosting are typically shallow, it would not be necessary to integrate conventional CART-style splitting, as in the UFO algorithm. The fact that the tree-root construction principle captures interaction effects without marginal effects better suggests that it could enhance the predictive performance of gradient boosting. This would be particularly noteworthy given that conventional gradient boosting with trees is already among the most powerful prediction methods, as evidenced by the widespread practical success of XGBoost \citep{Chen:2016}.

\myheading{Other Outcome Types}
While the UFO framework has been defined for both categorical and continuous outcomes, its performance has been evaluated only for categorical outcomes. The version for continuous outcomes, however, differs from that for categorical outcomes only in the calculation of the partition criterion and the split scores, where the reduction in variance is used instead of the reduction in Gini impurity. All other components of the framework remain identical, which is why the conclusions drawn from the analyses presented in this paper should largely extend to the version for continuous outcomes.

Beyond categorical and continuous outcomes, survival outcomes represent another important outcome type frequently encountered in practice, particularly in (bio)medical applications. Developing a variant of the UFO framework for survival outcomes would therefore be desirable. However, this would require defining a partition criterion suitable for survival outcomes. For categorical and continuous outcomes, a straightforward generalization of the reduction in Gini impurity and variance from the binary to the multiway case could be employed, as these criteria aim to maximize the purity of the outcome values within the subnodes. In contrast, the criterion most commonly used in survival trees---the log-rank test statistic---aims to maximize the differences in outcome distributions between child nodes. When dividing into more than two nodes, as required by the partition criterion in the UFO framework, this statistic can become large even if only one subnode differs markedly from the others, while the remaining subnodes are very similar. A suitable partition criterion for survival outcomes would, however, need to reward the homogeneity of outcome values within the end nodes, analogous to the criteria used for categorical and continuous outcomes.

\myheading{Limitations and Challenges}
Our simulation studies did not include high-dimensional data. This decision was mainly motivated by the fact that a key objective of the UFO framework is the interpretation of the strongest marginal or conditional effects using CRTRs. In high-dimensional settings, however, the curse of dimensionality makes the identification of individual interaction effects challenging and increases the risk of false positives. Moreover, the split points identified in the CRTRs may become less precise as the number of covariates considered per tree root increases. For these reasons, results obtained from applications of the UFO framework to high-dimensional datasets should be interpreted with caution.

In the simulation study, the unity VIM did not consistently rank informative covariates above non-informative ones for the smaller sample sizes considered, whereas the conventional VIMs tended to perform better in this setting. Although the exact reasons for the weaker performance of the unity VIM in small samples could not be clearly identified, it is likely related to the relatively high complexity of the UFO framework. Compared with conventional RFs and their associated VIMs, the UFO framework involves more computational steps and hyperparameters, which may contribute to the greater variability in the results. In addition, the simultaneous selection of all split covariates and split points during the construction of each tree root could lead to increased instability for small sample sizes---even though such instability was not observed for the related optimal trees in the simulation studies by \citet{Bertsimas:2017}. Nevertheless, it cannot be ruled out that the unity VIM may perform more reliably for small datasets with other DGPs than those used in our simulation study. For instance, when applied to the relatively small wine dataset ($n = 178$), the unity VIM produced results similar to those of the permutation VIM, which supports its applicability to smaller datasets. However, for such cases, it remains advisable to compare the results of the unity VIM with those of conventional VIMs to avoid misleading interpretations.

Weak effects were detected less reliably by the unity VIM than by conventional VIMs. This, however, does not represent a major limitation, as the primary aim of the unity VIM is to identify relevantly strong covariate effects---particularly those that conventional VIMs fail to capture---and to characterize the nature of these effects through CRTRs.

\myheading{Permutation VIM vs. Corrected Gini Importance for Interacting Covariates}
Another observation from the simulation study was that the permutation VIM better identified interacting covariates than the corrected Gini importance. A plausible explanation is that the corrected Gini importance reflects only the discriminatory power of splits on the covariate under consideration, whereas the permutation VIM---by permuting the covariate values---also affects the discriminatory power of splits in other covariates that interact with it.

This mechanism can be illustrated using a simple example. Consider a binary covariate $X_1 \in \{0,1\}$ that interacts with a continuous covariate $X_2$: when $X_1 = 0$, $X_2$ has no effect, but when $X_1 = 1$, it has a strong effect. In this case, it is likely that a subset $\mathcal{T}$ of the trees in the RF will first split on $X_1$, and then---due to the interaction---split on $X_2$ in the child node corresponding to $X_1 = 1$ (or in a subsequent node). When calculating the permutation VIM value of $X_1$, some observations with $X_1 = 1$ will, after permutation, be assigned the value $X_1 = 0$. In the trees from $\mathcal{T}$, this not only disrupts the discriminatory power of the split on $X_1$, but also removes the benefit of the subsequent split on $X_2$, since that split is no longer evaluated for these observations.

In general, similar mechanisms explain why covariates involved in interaction effects tend to receive higher permutation VIM values than corrected Gini importance values.

\section{Summary and Conclusions}
\label{sec:sum_concl}

The UFO framework is a data analysis approach for prediction problems with categorical and continuous outcomes. It enables a better detection and utilization of interaction effects without marginal effects while also providing interpretable insights into the strongest marginal or conditional effects of the covariates.

Default values for all hyperparameters were carefully chosen, and the suitability of the critical parameters was empirically assessed using 30 datasets. The influence of these parameters on predictive performance was found to be small.

In a large-scale comparative study involving 168 datasets, UFOs performed statistically significantly better than conventional RFs in terms of AUC and accuracy. For the Brier score, they outperformed RFs in slightly fewer than half of the datasets, although the difference was not statistically significant. Overall, the observed differences in predictive performance between the two methods were small.

In the simulation studies, the unity VIM reliably identified covariates exhibiting purely interaction-based effects, in contrast to the conventional measures---permutation importance and corrected Gini importance. Other covariate effects were consistently detected by all VIMs. Likely due to the higher algorithmic complexity of the UFO framework, the unity VIM was somewhat less reliable than conventional VIMs in distinguishing informative from non-informative covariates in small samples and for weak effects. For very small sample sizes, UFO results should therefore be interpreted with caution and compared with conventional methods.

The covariate-representative tree roots (CRTRs) reliably represented the true covariate effect types in simulated data. Furthermore, the application to a real dataset demonstrated that CRTRs can reveal covariate effects that would likely remain undetected in conventional analyses.

Overall, the UFO framework represents a powerful exploratory tool for identifying influential covariates and effect structures that are often overlooked by existing methods, thus offering the potential for new insights into dependency structures in the domain sciences.

\section*{Acknowledgements}
This work was supported by the German Science Foundation [DFG-Einzelförderung  HO 6422/1-3 to Roman Hornung].

\section*{{S}upplementary material and R code}
The supplementary material accompanying this article is available on the arXiv page. The R code used to produce the results shown in the main paper and in the supplementary material is available on GitHub
(\url{https://github.com/RomanHornung/UnityForests_code_and_data}, commit: 08edff9, accessed on January 11, 2026).

\bibliographystyle{abbrvnat}
\bibliography{foo}

\end{document}